\documentclass[fleqn,usenatbib]{mnras}

\usepackage{newtxtext,newtxmath}

\usepackage[T1]{fontenc}

\DeclareRobustCommand{\VAN}[3]{#2}
\let\VANthebibliography\thebibliography
\def\thebibliography{\DeclareRobustCommand{\VAN}[3]{##3}\VANthebibliography}

\usepackage{graphicx}	
\usepackage{amsmath}	
\usepackage{bm}
\usepackage{subcaption}
\usepackage{gensymb}
\usepackage{braket}
\usepackage[export]{adjustbox}

\newcommand*{\typewriter}{\fontfamily{lmtt}\selectfont{}}
\newcommand*{\taueq}{\tau_{\rm{eq}}}
\newcommand*{\tauin}{\tau_{\rm{in}}}
\newcommand*{\taudyn}{\tau_{\rm{dyn}}}
\newcommand*{\sigdyn}{\sigma_{\rm{dyn}}}
\newcommand*{\sigreg}{\sigma_{\rm{reg}}}
\newcommand*{\Msun}{\rm{M}_\odot}

\newcommand*{\Mstar}{\mathrm{M}_*}
\newcommand*{\sigmaMS}{\sigma_{\mathrm{MS}}}
\newcommand*{\qcr}{\fontfamily{qcr}\selectfont{}}

\newcommand*{\tage}{t_{\mathrm{age}}}
\newcommand*{\DeltaL}{\Delta_{\rm{L}}}
\newcommand*{\DeltaS}{\Delta_{\rm{S}}}
\newcommand*{\DeltaMS}{\Delta{\rm{MS}}}

\title[SFH variability in $z \sim 0.8$ galaxies]{Decoding the variability in the star-formation histories of $z \sim 0.8$ galaxies}

\author[Wan et al.]{
Jenny T. Wan,$^{1,2,3,4}$\thanks{E-mail: jtw48@cam.ac.uk}
Sandro Tacchella,$^{1,2}$
Francesco D'Eugenio,$^{1,2}$
Benjamin D. Johnson,$^{5}$
\newauthor
Arjen van der Wel$^{6}$
\\
$^{1}$Kavli Institute for Cosmology, University of Cambridge, Madingley Road, Cambridge, CB3 0HA, UK\\
$^{2}$Cavendish Laboratory, University of Cambridge, 19 JJ Thomson Avenue, Cambridge, CB3 0HE, UK\\
$^{3}$Department of Physics, Stanford University, 382 Via Pueblo Mall, Stanford, CA 94305, USA\\
$^{4}$Kavli Institute for Particle Astrophysics \& Cosmology, P.O. Box 2450, Stanford University, Stanford, CA 94305, USA\\
$^{5}$Center for Astrophysics | Harvard \& Smithsonian, 60 Garden St., Cambridge, MA 02138, USA\\
$^{6}$3 Sterrenkundig Observatorium, Universiteit Gent, Krijgslaan 281 S9, B-9000 Gent, Belgium
}

\date{Accepted XXX. Received YYY; in original form ZZZ}

\pubyear{2023}

\graphicspath{{figures/}}

\begin{document}
\label{firstpage}
\pagerange{\pageref{firstpage}--\pageref{lastpage}}
\maketitle

\begin{abstract}
The scatter of the star-forming main sequence (SFMS) holds a wealth of information about how galaxies evolve. The timescales encoded in this scatter can provide valuable insight into the relative importance of the physical processes regulating star formation. In this paper, we present a detailed observational analysis of the timescales imprinted in galaxy star-formation history (SFH) fluctuations by using the stochastic SFH model to fit 1928 massive, $z\sim0.8$ galaxies in the LEGA-C survey. We find that the total intrinsic scatter of the SFMS is $\sim 0.3$ dex in galaxies with stellar masses $\gtrsim 10^{10}~\Msun$. This scatter decreases as the timescale over which SFRs are averaged increases, declining to a non-negligible $\sim 0.15-0.25$ dex at 2 Gyr, underscoring the importance of long-timescale SFH diversity to the SFMS scatter. Furthermore, galaxies currently above (below) the SFMS tend to have been above (below) the SFMS for at least $\sim 1$ Gyr, providing evidence that individual galaxies may follow different median tracks through SFR$-\Mstar$ space. On shorter timescales ($\sim 30 - 100$ Myr), galaxies' SFRs also vary on the order of $\sim 0.1-0.2$ dex. Our work supports the idea that the SFMS emerges from a population average of the pathways that individual galaxies trace through the SFR$-\Mstar$ plane. The scatter reflects the long-term heterogeneity of these paths likely set by the evolutionary timescales of halo growth and cooling, accentuated by short-term variations reflecting the dynamical timescale of the galaxy and its interstellar medium. Our results emphasize the dynamic nature of the SFMS and the importance of understanding the diverse processes governing star formation.
\end{abstract}



\section{Introduction}

The large galaxy surveys of the past few decades 
have thrown open the blinds to reveal just how varied galaxies in the observable universe truly are. But despite this incredible diversity, trends and scaling relations have been established across the galaxy population. 

One such scaling law is the star-forming main sequence (SFMS), which describes the correlation between the stellar mass ($\Mstar$) and the star formation rate (SFR) in star-forming galaxies that exists from present day up to at least redshift $z \sim 6$ \citep{Brinchmann2004, Daddi2007, Elbaz2007, Whitaker2012, Speagle2014, Schreiber2015, Boogaard2018, Leja2022, Popesso2023}. The specific properties of this relation have been studied extensively in the preceding decade. The exact shape of the SFMS is still under debate, but many studies point to a power law, SFR $\propto \Mstar^\alpha$ (in the local universe, e.g., \citealt{Peng2010, RenziniPeng2015}; and in the early universe, e.g., \citealt{Speagle2014, Pearson2018}). Others find evidence of a bending in the SFMS shape at the high-mass end \citep[stellar masses above $\sim10^{10.5}$ M$_\odot$; e.g.,][]{Schreiber2015, Whitaker2015, Popesso2019, Leja2022, Stephenson2024}. Additionally, the normalization of the SFMS is observed to increase with lookback time (or redshift, $z$) and is often expressed as $(1 + z)^\gamma$, with $\gamma$ varying between 1.9 and 3.7 \citep{Speagle2014, Whitaker2014, Schreiber2015, Leja2022}. This trend is a result of the higher dark matter halo accretion rates in the early (high-$z$) universe, which lead to higher gas fractions and larger overall SFRs in high-$z$ galaxies \citep{Bouche2010, Lilly2013, Tacchella2018}. 

Perhaps the most interesting feature of the SFMS, however, is its observed scatter of $\sigmaMS \sim 0.2-0.4$ dex at any given redshift. A key implication of this scatter is that galaxies along the main sequence ridgeline self-regulate their star formation through a dynamic equilibrium cycle of gas inflow, outflow and consumption \citep{Lilly2013, Rodriguez2016, Tacchella2016, Matthee2019}. 

Based on the VELA \citep{Ceverino2014, Zolotov2015} cosmological zoom-in simulations, \citet{Tacchella2016} suggested that galaxies are confined to a narrow $\pm0.3$~dex envelope around the SFMS and can move into the upper envelope of the SFMS via gas compaction resulting from, e.g., mergers or disk instabilities. As their central gas reservoirs become depleted by active star formation and outflows, these systems will eventually fall back onto --- or below --- the SFMS. Their SFRs can increase yet again if inflows are able to effectively replenish the gas supply and trigger a new round of compaction. This up-and-down movement of galaxies about the SFMS manifests as the observed $\sigmaMS$. \citet{Rodriguez2016} argued, using the $N$-body \textit{Bolshoi-Planck} simulation \citep{Klypin2016}, that the scatter of the SFMS could be set by the $\sim 0.3$ dex dispersion found in the halo mass accretion rate. With the EAGLE simulations \citep{Crain2015, Schaye2015}, \citet{Matthee2019} showed that locally, the SFMS scatter originates from a combination of self-regulation-associated short-term ($0.2 - 2$ Gyr) fluctuations and halo formation-associated longer-term ($\sim 10$ Gyr) fluctuations. Additionally, using the THESAN-ZOOM simulations, \citet{McClymont2025} observed an increase in the overall scatter of the SFMS with cosmic time, driven by long-term environmental effects, as well as an increase in the short-term SFMS scatter at high redshifts where star formation is burstier.

While there is evidence for the SFMS as an evolutionary sequence along which galaxies oscillate as they grow their stellar masses, it is not clear that this interpretation is necessarily the correct one. In fact, ``physics-free'' models of galaxy evolution have had remarkable success in recovering several important observables. By approximating stellar mass growth as a stochastic process without any prior physical inputs, \citet{Kelson2014} and \citet{Kelson2016} were able to match the observed SFMS over $0 \leq z \leq 10$, the scatter in specific star-formation rates (sSFRs) at fixed mass, and the Tully-Fisher relation, as well as the shape and evolution of the stellar mass function over $2 < z < 12.5$. Additionally, using nothing more than loosely-constrained log-normals to model galaxy SFHs, \citet{Abramson2016} faithfully recovered the SFMS slope at $z \leq 6$, the stellar mass functions at $z \leq 8$, and the rapid and gradual channels of galaxy quenching. In these somewhat extreme scenarios, the SFMS is not a semi-deterministic relationship indicating that galaxies concurrently grow in mass and increase their SFRs, but an emergent property of the diverse, independent SFH trajectories traced by galaxies over time \citep{Gladders2013, Kelson2014, Abramson2016}. 

It is, perhaps, slightly disconcerting that models involving very limited physics are able to produce key scaling relations to similar levels of fidelity as complex cosmological simulations. Does this mean that the paths along which galaxies grow over time are completely independent of the physical processes at work within them? It is likely not so dramatic as that. Observations and simulations alike have demonstrated that AGN and supernovae feedback play a role in modulating the interstellar medium within galaxies \citep[e.g.,][]{Henriques2019, Mulcahey2022}; up to $ z\sim 8$, we have observed powerful outflows driving gas and metals out of galaxies  \citep[e.g.,][]{Bischetti2019, Gallagher2019, Carninani2023}; we see galaxy-galaxy interactions occurring from the local universe out to redshifts of $z \gtrsim 4-10$ \citep[e.g.,][]{Barton2000, Cibinel2019, Hsiao2023, Suess2023}. These processes do not work in isolation without in some way affecting the galactic environments they live in. 

On the other hand, studies of the clustering properties of the galaxy population from $z \sim 0.7$ down to $z \sim 0$ have demonstrated that galaxy clustering depends just as --- if not more --- strongly on specific star-formation rate (sSFR $=$ SFR$/\Mstar$) as on stellar mass \citep{Coil2017, Berti2021}. The presence of distinct clustering properties for galaxies above and below the SFMS indicates that the scatter observed in the main sequence corresponds to a physical property of the larger-scale environment. 

The question then becomes, how important are these different mechanisms in driving the evolution of galaxies? In other words, does the dominant physics lie in the processes which regulate star-formation in individual galaxies, or in the processes that diversify the star-formation histories of the galaxy population? Do galaxies evolve \textit{along} the SFMS, or \textit{across} it, on their road to quiescence? 

Key information about the growth and evolution of galaxies is hidden within the scatter of the SFMS. The SFMS dispersion relates directly to the variability in the star-formation histories (SFHs) of individual galaxies, and thus encodes the physical processes --- and associated timescales --- which regulate star formation. If galaxies follow smooth, log-normal growth histories \citep{Gladders2013,Abramson2016}, $\sigmaMS$ predominantly reflects the Hubble-timescale diversification of galaxies. If star formation is a purely stochastic, random-walk-like process \citep{Kelson2014}, $\sigmaMS$ consists of fluctuations on arbitrary timescales and is physically meaningless. If galaxy evolution follows the grow-and-quench prescription of cosmological simulations, the fluctuations in the SFRs of galaxies are tied to mechanisms ranging from the local creation and destruction of giant molecular clouds (GMCs) on short timescales \citep[$\lesssim$ 100 Myr;][]{Scalo1984, Krumholz2015, Orr2019}, to galaxy-galaxy mergers and galactic winds from stellar and AGN feedback on intermediate timescales \citep[$\sim 0.1 - 1$ Gyr;][]{GunnGott1972, Hernquist1989, WangLilly2020}, to galactic outflows, dark matter halo accretion rates, and environmental large-scale structure on the longest timescales \citep[$\gtrsim$ 1 Gyr;][]{Rodriguez2016, Tacchella2018, Behroozi2019}. The relative importance of these different processes to galaxy growth can be constrained by analyzing the variability in the SFHs of galaxies across a wide range of timescales \citep{Iyer2020, TFC2020, Shin2023}.

Assuming that a galaxy's SFH is a stochastic process, the relative importance of different timescales can be described by its power spectral density \citep[PSD;][]{CT2019, TFC2020, Iyer2020, Iyer2024}. The PSD effectively quantifies the amount of power contained in SFR fluctuations (i.e., burstiness) as a function of timescale in the SFHs of a galaxy population. \citet{Wan2024} built upon this PSD framework to develop a physically-motivated, ``stochastic'' prior for non-parametric SFHs in the
spectral energy distribution (SED)-modelling code {\typewriter Prospector}. 

By applying the stochastic SFH prior to a high-resolution sample of $z \sim 0.8$ galaxies observed with the LEGA-C survey \citep{LEGAC2021}, we aim to quantify the way in which galaxies evolve about the SFMS. Many previous works have investigated the stellar population properties and SFHs of LEGA-C galaxies \citep[e.g.][]{Wu2018, Sobral2022, Cappellari2023, Bevacqua2024, Kaushal2024, Steel2024, Nersesian2025}. New to our work is the quantitative consideration of the questions, \textit{What timescales are important in the SFHs of galaxies?} and \textit{What can the variability in their SFHs tell us about the physical processes that regulate their growth and evolution?} Here, we present the detailed observational constraints on parameters corresponding to SFH variability in massive galaxies at $z = 0.6 - 1.0$.

The paper is structured as follows. Section \ref{sec:data} describes the observational data and galaxy sample used in our analysis. Section \ref{sec:spectro-photometric modelling} details the physical model adopted in the SED-modelling of our observational data. We present some of the key reconstructed galaxy properties in Section \ref{sec:reconstructed galaxy properties}. In Section \ref{sec:characterizing the SFMS}, we characterize the dispersion of galaxies on the SFMS, as well as the relative importance of long- versus short-timescale SFH variability, and discuss the implications and avenues for future work in Section \ref{sec:discussion}. We conclude in Section \ref{sec:conclusions}. 

Throughout this work, we assume a flat $\Lambda$CDM cosmology with $H_0 = 70 ~\rm{km ~s^{-1} Mpc^{-1}}$, $\Omega_m = 0.3$, and $\Omega_\Lambda = 0.7$.

\section{Observational data}
\label{sec:data}
In this section, we describe the spectroscopic and photometric data used in our analysis, as well as the selection criteria implemented to arrive at our final sample.

\subsection{Spectroscopy}
\label{sec:Spec}
The spectroscopic data are obtained from the third data release (DR3) of the Large Early Galaxy Astrophysics Census \citep[LEGA-C,][]{LEGAC2021}. LEGA-C is a public spectroscopic survey targeting $0.6 \lesssim z \lesssim 1.0$ galaxies in the COSMOS field. DR3 contains high signal-to-noise (S/N), high resolution spectra for 4081 galaxies, selected from the UltraVISTA catalog \citep{Muzzin2013} using a redshift-dependent $K_s$ magnitude limit, which ranges from $K_{s} \approx 21.1 - 20.4$. The spectra were collected with ESO's Very Large Telescope/Visible Multi-Object Spectrograph \citep[VIMOS,][]{VIMOS2003} with integration times of $\sim 20$ hours. 

\subsection{Photometry}
\label{sec:Phot}
The accompanying photometric data are provided by the COSMOS2020 photometric catalog \citep{COSMOS2020} and the COSMOS Super-deblended catalog \citep{Superdeblended}. The COSMOS2020 catalog contains $\sim 1$ million sources observed in 32 bands from the (rest) UV to near-IR. The Super-deblended COSMOS catalog presents photometry for $\sim11,000$ galaxies over the mid- to far-IR/(sub)mm ($100 - 1200~\mu\mathrm{m}$) range in 11 filters.

\subsection{Galaxy sample}
\label{sec:galaxy sample}

We begin with the full sample of galaxies from the LEGA-C DR3 catalog. We require the galaxies to be primary targets of the LEGA-C survey ({\qcr PRIMARY} = 1) and to lie within a redshift range of $0.6 \leq z \leq 1.0$ and a nominal stellar mass range of $10.3 \leq \log \mathrm{M_*}/\Msun \leq 11.5$.
Furthermore, we implement a S/N floor of $10~\Angstrom^{-1}$ (measured at rest-frame $4000\Angstrom$) so that the resolution of the spectra are sufficient for extracting SFR timescale information using the stochastic SFH prior. 
Lastly, for simplicity of modelling, we eliminate galaxies that show evidence of AGNs and disturbed morphologies (e.g., are mergers) by ensuring that {\qcr FLAG\_SPEC} = 0 and {\qcr FLAG\_MORPH} = 0. This results in a working sample of 1973 galaxies.

\begin{figure}
    \centering
    \includegraphics[width=0.48\textwidth]{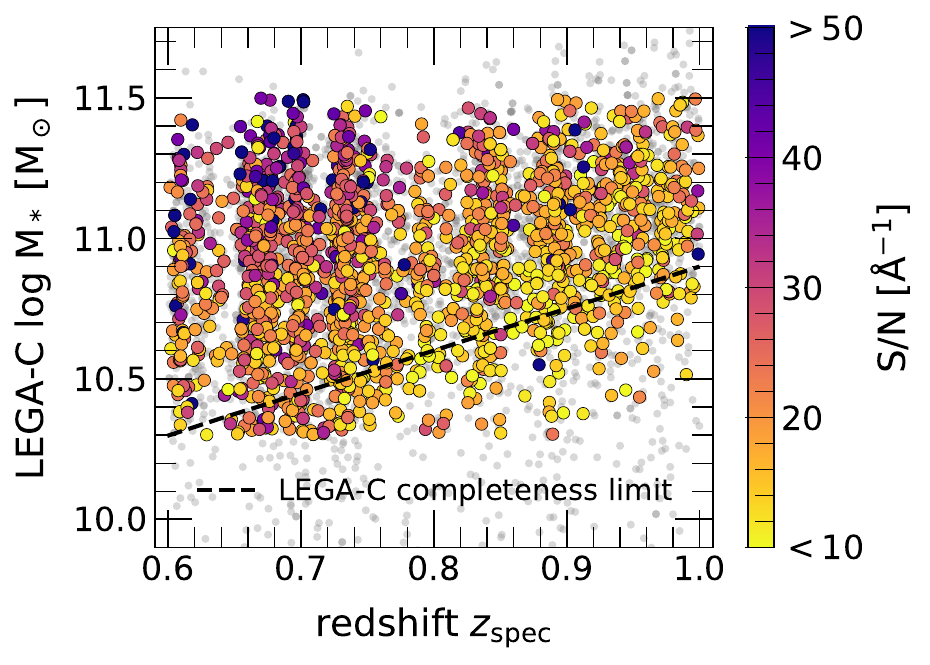}
    \caption{Our galaxy sample in the $\log\Mstar-z_\mathrm{spec}$ plane, color-coded by S/N measured at $4000\Angstrom$ (all three quantities are reported from the LEGA-C DR3 catalog). The gray points represent the underlying galaxy distribution from the COSMOS2020 catalog. The black dashed line marks the mass-completeness threshold of the LEGA-C survey \citep{LEGAC2021}. We select galaxies within a redshift range of $0.6 \leq z \leq 1.0$ and a stellar mass range of $10.3 \leq \log \mathrm{M_*/M_\odot} \leq 11.5$ for this analysis. Our final sample of 1928 galaxies have spectra with S/N between $10.0~\Angstrom^{-1}$ and $97.2~\Angstrom^{-1}$, with an average S/N of $21.6~\Angstrom^{-1}$. The average redshift of our sample is $\braket{z_{\mathrm{spec}}} = 0.78$.}
    \label{fig:sample distribution}
\end{figure}

From this subsample, we select galaxies which have matches in both the COSMOS2020 catalog and the Super-deblended catalog. These selection criteria result in a final sample of 1928 galaxies. Figure \ref{fig:sample distribution} shows the LEGA-C DR3 measurements of stellar mass ($\log \Mstar$) as a function of observed spectroscopic redshift ($z_{\mathrm{spec}}$) for this sample. The mass-completeness threshold of the LEGA-C survey as a function of redshift is marked with a black dashed line. The colors of the points correspond to the S/N of the spectra measured at $4000\Angstrom$. The core of our sample lies in the $10.7 < \log \Mstar/\Msun < 11.3$ range, where there is a fairly uniform redshift distribution ($\braket{z_{\mathrm{spec}}} = 0.8$). There is no distinct trend in S/N; however, we do find that galaxies at $z_{\mathrm{spec}} < 0.7$ have, on average, a slightly higher S/N than those at $z_{\mathrm{spec}} > 0.7$. Additionally, while the LEGA-C survey is mass-complete down to $\sim10^{10.3}~\Msun$ at $z \sim 0.6$, this threshold increases to $\sim10^{10.9}$ at $z \sim 1$ \citep{LEGAC2021}. Thus, this sample is likely missing lower-mass galaxies at $z \gtrsim 0.8$.


\section{Spectro-photometric modelling}
\label{sec:spectro-photometric modelling}

In this section, we introduce the physical model and priors used to model the observational data described in the previous section (Section \ref{sec:model and priors}). We present the fitting results in Section \ref{sec:fitting results} and verify that the adopted model provides a good fit to the spectral and photometric data.

\begin{table*}
    \centering
    \caption{Description of the free parameters and corresponding priors used in the {\typewriter Prospector} fits. Our adopted model contains 32 free parameters.}
    \label{tab:priors}
    \begin{tabular}{lll}
        \hline
        \hline
        Parameter & Description & Prior \\
        \hline
        $z$ & Redshift & Uniform~($z_{\mathrm{spec}} - 0.005$, $z_{\mathrm{spec}} + 0.005$), where $z_{\mathrm{spec}}$ is the spectroscopic \\

& & \hspace{1em} redshift \\

$\sigma_{*}$/(km s$^{-1}$) & Stellar velocity dispersion & Uniform~(40.0, 400.0) \\ 

log(M$_*$ / M$_\odot$) & Stellar mass & Uniform~(9.5, 12.0) \\

log(Z$_*$ / Z$_\odot$) & Stellar metallicity & Uniform~($-1.0$, 0.19) \\

log SFR ratios & Ratio of SFRs in adjacent time bins & Multivariate-Normal~($\bm{\mu} = 0$, $\Sigma = \mathrm{ACF}_{\mathrm{SFH}}(\sigma, \taueq, \tauin, \sigdyn, \taudyn)$) \\

$\sigma_{\mathrm{reg}}/\mathrm{dex}$ & Overall variability in gas inflow and cycling processes & log-Uniform~(0.1, 5.0) \\

$\tau_{\rm{eq}}$ / Gyr & Equilibrium timescale & Uniform~(0.01, $t_H$), where $t_H$ is the age of the universe at the redshift of \\

& & \hspace{1em} the object \\

$\sigma_{\rm{dyn}}/\mathrm{dex}$ & Overall variability in short-timescale dynamical processes & log-Uniform~(0.001, 0.1) \\

$\tau_{\rm{dyn}}$ / Gyr & Dynamical timescale & Clipped-Normal~(min $= 0.005$, max $= 0.2$, $\mu= 0.01$, $\sigma = 0.02$) \\

$n$ & Power-law multiplicative modifier to Calzetti law & Uniform~($-1.0$, 1.0) \\

$\hat{\tau}_{\rm{dust, 2}}$ & Diffuse dust optical depth & Clipped-Normal~(min $=0.0$, max $=4.0$, $\mu=0.3$, $\sigma=1$) \\

$\hat{\tau}_{\rm{dust, 1}}$ & Birth cloud optical depth & Clipped-Normal in ($\hat{\tau}_{\rm{dust, 1}}$ / $\hat{\tau}_{\rm{dust, 2}}$) (min$=0.0$, max$=2.0$, $\mu=1.0$, \\
& & \hspace{1em}$\sigma=0.3$) \\

$U_{\rm{min}}$ & Minimum radiation field strength & Clipped-Normal~(min$=0.1$, max$=15.0$, $\mu=2.0$, $\sigma=1.0$) \\

$\gamma_e$ & Fraction of dust heated at radiation intensity $U_{\rm{min}}$ & log-Uniform~($10^{-4}$, $0.1$) \\

$q_{\rm{PAH}}$ & Fraction of dust grain mass in PAHs & Uniform~(0.5, 7.0) \\

$\sigma_{\rm{gas}}$ & Velocity dispersion of gas & Uniform~($30.0$, $300.0$) \\

$\log(\mathrm{Z}/\mathrm{Z_\odot})$ & Gas-phase metallicity & Uniform~($-2.0$, $0.5$) \\

$\log(U)$ & Gas ionization parameter for nebular emission & Uniform~($-4.0$, $-1.0$) \\

$f_{\rm{out}}$ & Fraction of spectral data points considered outliers & Uniform~($10^{-5}$, $0.5$)  \\
        \hline
    \end{tabular}
\end{table*}

\subsection{SED-fitting model and priors}
\label{sec:model and priors}

We use the SED-modeling code {\typewriter Prospector} \citep{JohnsonLeja2017, Johnson2021} to simultaneously fit the photometric and spectroscopic data for the galaxies in our sample. The {\typewriter dynesty} dynamic nested sampling package \citep{Speagle2020} is used to sample the posterior probability distribution. Table \ref{tab:priors} summarizes the free parameters and associated priors used in our physical model. We describe the various model components in further detail below.

\subsubsection{Stellar population model}

Stellar population synthesis in {\typewriter Prospector} is handled by the Flexible Stellar Population Synthesis ({\typewriter FSPS}) package \citep{Conroy2009, ConroyGunn2010}, accessed through {\typewriter PYTHON-FSPS}\footnote{\url{https://github.com/dfm/python-FSPS}} \citep{ForemanMackey2014}. We use {\typewriter FSPS} configured with the MIST isochrones \citep{Dotter2016, Choi2016} and the empirical MILES spectral library \citep{MILES2011} to conduct stellar population synthesis. The MILES library consists of $\sim$~1000 stellar spectra covering a wavelength range of $3525 - 7500 \Angstrom$ at a resolution of $2.5 \Angstrom$ (FWHM). The Modules for Experiments in Stellar Astrophysics (MESA) code \citep{Paxton2011, Paxton2013, Paxton2015} is used to compute the stellar evolutionary tracks from which the MIST isochrones are constructed.

We allow for slight deviations from the spectroscopic redshift reported in the LEGA-C DR3 catalog by setting a tight prior on the redshift, allowing it to vary uniformly within $\pm 0.005$ of the catalog $z_{\rm{spec}}$. The prior on stellar mass is uniform between $9.5 \leq \rm{log~M_*/M_\odot} \leq 12.0$, and the prior on stellar velocity dispersion is uniform between $40 \leq \sigma_*/\rm{km~s^{-1}} \leq 400$. Additionally, we assume that all stars within a galaxy share the same metal content. This single metallicity, reported as $\log(\mathrm{Z_* / Z_\odot})$ with $\mathrm{Z_\odot} = 0.0142$, is allowed to vary within a uniform prior between $-1.0$ and $0.19$. The upper limit of the stellar metallicity prior is determined by the metallicities for which there is adequate coverage of the HR diagram in the MILES stellar templates. Lastly, we adopt the \citet{Kroupa2001} initial mass function throughout this work.

\subsubsection{Star formation history}
\label{sec:sfh}

We describe the star-formation activity of galaxies in our sample with a non-parametric SFH model, which fits for the log of the ratio between SFRs in adjacent time bins, $\log \mathrm{SFR~ratios} \equiv \big\{\log( \mathrm{SFR}_n/\mathrm{SFR}_{n+1}) \big\}$, where $n = 0, 1, \dots, N-1$ for a given number of SFH bins $N$. We apply the stochastic SFH prior described in \citet{Wan2024}, meaning a multivariate normal prior is placed on $\log\mathrm{SFR~ratios}$, with a mean of zero and a covariance matrix determined by $\sigreg$, $\taueq$, $\tauin$, $\sigdyn$, and $\taudyn$ (the ``PSD parameters'').

Section 2.2.2 of \citet{Wan2024} describes in detail how this covariance matrix is defined. Here, we provide some basic intuition for the meaning of the individual PSD parameters. $\taueq$, $\tauin$, and $\taudyn$ are the effective timescales associated with equilibrium gas cycling (between atomic and molecular gas), changes to the gas reservoir (e.g., inflows), and dynamical processes that modulate the life cycles of GMCs, respectively; $\sigreg$ encodes the overall variability due to long-timescale gas reservoir-regulating processes; and $\sigdyn$ describes the variability due to shorter-timescale GMC-regulating processes. Together, these parameters set the correlation structure of the stochastic star-formation process.

\citet{Wan2024} performed a series of recovery tests on mock LEGA-C data using the stochastic SFH model and demonstrated that the stochastic prior is able to infer both basic stellar population parameters (e.g., stellar mass, stellar metallicity, dust index), as well as higher-order properties like SFR and mass-weighted age, to a high level of accuracy with very minimal bias. Thus, we adopt the same priors on the PSD parameters as used in \citet{Wan2024}, allowing $\sigreg$, $\taueq$, $\sigdyn$, and $\taudyn$ to vary within the priors listed in Table \ref{tab:priors} and fixing $\tauin$ to $t_H$ (the Hubble time at the epoch of observation) for each galaxy. (Gas inflow rates are correlated with the accretion history of galaxies' parent halos, which evolve over timescales comparable to the age of the universe.) As shown in the aforementioned work, this prevents the SFH model from having more freedom than the data is equipped to handle.

\citet{Wan2024} additionally revealed that, to first order, $\sigreg$ controls the width of the SFH prior. The other PSD parameters act simply as perturbations on top of this. Therefore, $\sigreg$ is the only parameter for which reliable constraints can routinely be derived from our data. Thus, we do not discuss estimates of $\taueq$, $\sigdyn$, and $\taudyn$ in our analysis, as we simply marginalize over their uncertainties in the SED-fitting procedure.

The number of SFH bins is set to 14 in our analysis. The time bins are specified in look-back time, with the first two bins fixed at ($1 - 5$) Myr and ($5 - 10$) Myr, and the last bin ending at $0.95 t_H$ Gyr. The remaining bins are equally spaced in logarithmic time between $10$ Myr and $0.95 t_H$ Gyr.

\subsubsection{Dust attenuation \& emission}
We adopt a two-component dust attenuation model based on \citet{CharlotFall2000}, which considers the attenuation of young and old stellar light separately. Specifically, young stars (i.e., stars formed within the last 10 Myr) are expected to still be embedded within their ``birth clouds'', which results in stronger attenuation of their nebular and stellar emission than that of older stellar populations. This birth-cloud attenuation component, which only applies to stellar populations $\leq 10$ Myr old, follows the form:
\begin{equation}
    \tau_{\lambda,1} = \hat{\tau}_{\mathrm{dust,1}} \Big{(} \frac{\lambda}{5500 \Angstrom} \Big{)}^{-1}.
\end{equation}
The attenuation of all stellar and nebular emission from the galaxy is described by the diffuse component, which follows:
\begin{equation}
    \tau_{\lambda,2} = \frac{\hat{\tau}_{\mathrm{dust,2}}}{4.05} [k'(\lambda) + D(\lambda)] \Big{(} \frac{\lambda}{5500 \Angstrom} \Big{)}^n.
\end{equation}
$\hat{\tau}_{\mathrm{dust,1}}$ and $\hat{\tau}_{\mathrm{dust,2}}$ are the optical depths of the birth clouds and diffuse dust, respectively, $k'(\lambda)$ is the \citet{Calzetti2000} attenuation curve, and $D(\lambda)$ is the Drude profile which describes the UV dust bump. Following the results of \citet{KriekConroy2013}, the strength of UV bump and the offset in slope from Calzetti are tied to the slope of the diffuse dust attenuation curve, $n$. We assume a flat prior on $n$, letting its value vary uniformly between $-1$ and $1$. An informative, clipped-normal prior is adopted for $\hat{\tau}_{\mathrm{dust,2}}$.

Additionally, despite the fact that they are often degenerate, it is important to distinguish between $\hat{\tau}_{\mathrm{dust,1}}$ and $\hat{\tau}_{\mathrm{dust,2}}$ to accurately predict emission line properties. The optical depth towards nebular emission lines (generally from the ionizing radiation of young, hot stars) is $\sim 2 \times$ that of the stellar continuum \citep{Calzetti1994, Price2014}. This implies that $\hat{\tau}_{\mathrm{dust,1}} \sim \hat{\tau}_{\mathrm{dust,2}}$, as $\hat{\tau}_{\mathrm{dust,2}}$ applies to the entire galaxy and $\hat{\tau}_{\mathrm{dust,1}}$ applies only to the young stellar populations. We place an informative joint prior on the ratio $\hat{\tau}_{\mathrm{dust,1}}/\hat{\tau}_{\mathrm{dust,2}}$, a clipped-normal distribution centered at 1.0 with a width of 0.3 in the range $0.0 < \hat{\tau}_{\mathrm{dust,1}}/\hat{\tau}_{\mathrm{dust,2}} < 2.0$.

The dust emission from galaxies is modeled with the assumption that all starlight attenuated by dust is re-emitted in the IR. The shape of the IR SED is then described by the IR emission spectra from \citet{DraineLi2007}. These dust emission templates follow the silicate-graphite-PAH model \citep{Mathis1977, DraineLee1984}, where interstellar dust is taken to be a mixture of amorphous silicate and carbonaceous grains, including varying amounts of polycyclic aromatic hydrocarbon (PAH) particles.

In the \citet{DraineLi2007} model, the IR emission depends on three free parameters:
\begin{enumerate}
    \item $U_{\rm{min}}$, the minimum radiation field strength to which the dust mass is exposed;
    \item $\gamma_e$, the fraction of dust mass exposed to this minimum radiation field intensity;
    \item $q_{\rm{PAH}}$, the fraction of the total dust mass that is in PAHs. 
\end{enumerate}
$U_{\rm{min}}$ and $\gamma_e$ dictate the shape and location of the thermal dust emission bump in the IR SED, and $q_{\rm{PAH}}$ controls much of the emission in the MIR. We adopt informative priors on $U_{\rm{min}}$ and $\gamma_e$, and a flat prior on $q_{\rm{PAH}}$ (see Table \ref{tab:priors}).

\subsubsection{Nebular emission}
The nebular emission model in {\typewriter FSPS} assumes that the ionizing continuum from the model stellar populations is completely absorbed by the gas and re-emitted as both line and continuum emission. This emission is generated using the {\typewriter CLOUDY} \citep{Ferland1998, Ferland2013} grid implemented within {\typewriter FSPS} \citep[for details, see][]{Byler2017}. We adopt flat priors on the gas-phase metallicity ($-2.0 < \log(\mathrm{Z}/\mathrm{Z_\odot}) < 0.5$ and the gas ionization parameter ($-4.0 < \log(U) < 1.0$). 

However, because emission lines in quiescent and transitioning galaxies are tied to processes unrelated to star-formation (e.g., LINERs, shocks, AGN), we implement a flexible approach to modelling emission lines in these cases. As described in Appendix E of \citet{Johnson2021}, each emission line is modeled as a Gaussian with a variable width ($\sigma_{\rm{gas}}$, the velocity dispersion of the gas) and amplitude. We fit for $\sigma_{\rm{gas}}$, while the emission line amplitudes are marginalized over in each fitting step, with a prior based on the {\typewriter CLOUDY-FSPS} predictions. A uniform prior is applied to the gas-phase velocity dispersion ($30$ km s$^{-1}$ $< \sigma_{\rm{gas}} < 300$ km s$^{-1}$).

\subsubsection{Nuisance parameters}
\label{sec:nuisance params}
To account for any continuum mismatch issues, we exclude the continuum shape from our inference of the physical parameters. At each likelihood call, we multiply the model spectrum by a order 10 Chebyshev polynomial calibration function, parameterized as a function of wavelength, to produce the observed spectrum. We find that our results are generally insensitive to changes in the order of the polynomial. The least-squares maximum likelihood fit for the calibration is used in order to minimize computational cost. This spectroscopic calibration approach effectively pins the model continuum shape and normalization solely on the photometry such that the spectral continuum does not inform any of the galaxy's physical parameters.

We mitigate the effects of outlying spectroscopic data points, i.e., spectral data which are not well-described by the model (either due to underestimated uncertainties or model limitations), by including an outlier model. {\typewriter Prospector}'s outlier model follows the mixture model approach of \citet{Hogg2010}. No individual pixels are identified as outliers; rather, the likelihood function is modified by assuming some probability $f_{\rm{out}}$ that any given spectral pixel is an outlier and marginalizing over $f_{\rm{out}}$ for each pixel. We inflate the uncertainties of the outlying pixels by a factor of 50. Thus, the likelihood equation for spectroscopic fitting becomes $\mathcal{L}_{\rm{spec}} = (1 - f_{\rm{out}}) \mathcal{L}(f, m, \sigma) + f_{\rm{out}} \mathcal{L}(f, m, 50\sigma)$, where $f$ and $m$ are the observed and model fluxes, respectively (see Section 4.4 of \citealt{Tacchella2022Halo7D}). We include $f_{\rm{out}}$ as a free parameter in the model, and find that it is typically less than 0.01 in our fits.


\subsection{Fitting results}
\label{sec:fitting results}

\begin{figure}
     \centering
     \includegraphics[width=0.48\textwidth]{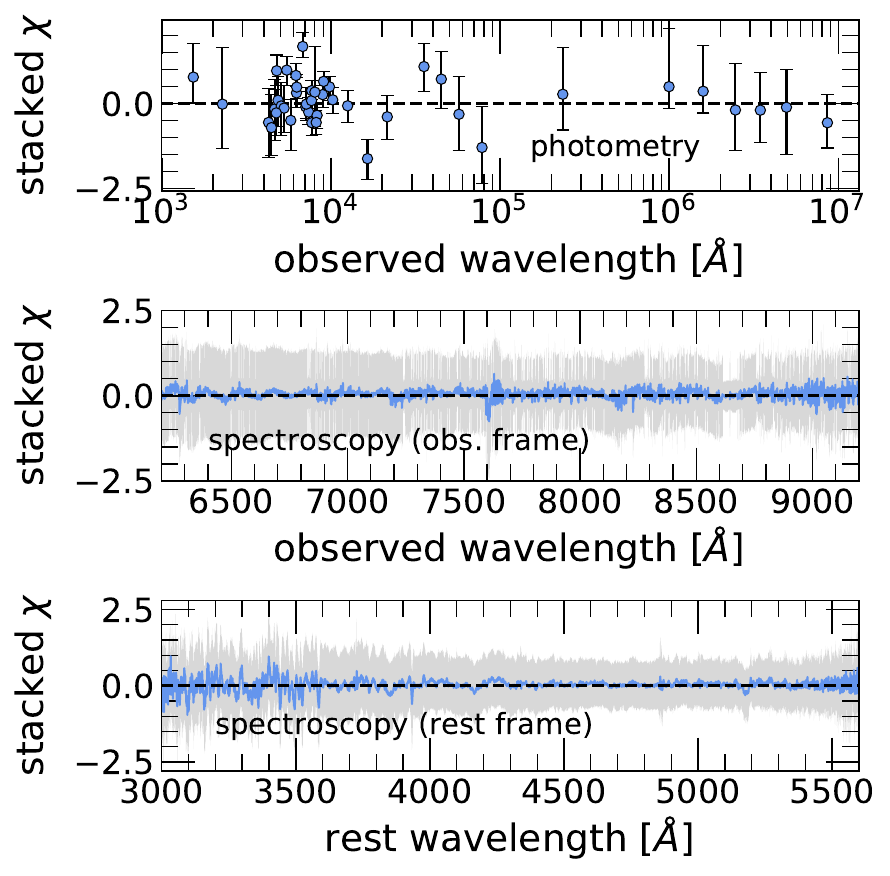}
    \caption{The stacked $\chi$ values of the {\typewriter Prospector} fits to the galaxies in our sample. The top panel shows the stacked $\chi$ for the photometry; the middle panel shows the stacked $\chi$ of the spectra in the observed frame; and the bottom panel is the stacked $\chi$ of the rest-frame spectra. The gray shaded regions in the middle and lower panels show the 16th$-$84th percentiles. We find that overall, the model reproduces the data well. The spectroscopic stacked $\chi$ values in both the rest and observed frames are consistent with zero. Specifically, the rest-frame shows that our model is able to reproduce all key spectral features; the observed-frame shows no calibration (by construction) or skyline issues. The majority of the photometric $\chi$ values are also consistent with zero.}
    \label{fig:stacked chis}
\end{figure}

We verify the goodness of the fits to the photometry and spectroscopy across the entire sample of galaxies. Figure \ref{fig:stacked chis} shows the stacked $\chi$ values for the photometric fits (top panel) and the spectroscopic fits in the observed and rest frames (middle and bottom panels, respectively), where $\chi$ = (the observed data $-$ the model data)/observational uncertainty. The observed-frame spectroscopic residuals trace potential instrument and/or observational issues, while the rest-frame traces potential physical inconsistencies (e.g., in the emission line modeling).

In general, the model reproduces the data well. The spectroscopic stacked $\chi$'s in both the rest and observed frames are consistent with zero. In particular, the goodness-of-fit in the rest-frame shows that our model is able to reproduce all key spectral features well. The observed-frame verifies that we are subject to neither calibration nor skyline issues. The majority of the photometric $\chi$'s are also consistent with zero. A number of the Subaru Suprime-Cam and Hyper Suprime-Cam bands, as well as the IRAC band 1 and the VISTA VIRCAM $H$-band, have $\chi$ values which are somewhat offset. Most notably, the Subaru Suprime-Cam IA679 band has a $\chi$ of about 1.7, which potentially calls for a zero-point correction in the future. However, the distribution of the photometric stacked $\chi$ values is scattered around zero, meaning that there is no overall bias in the photometric fits.

\begin{figure*}
    \centering
    \includegraphics[width=0.745\textwidth]{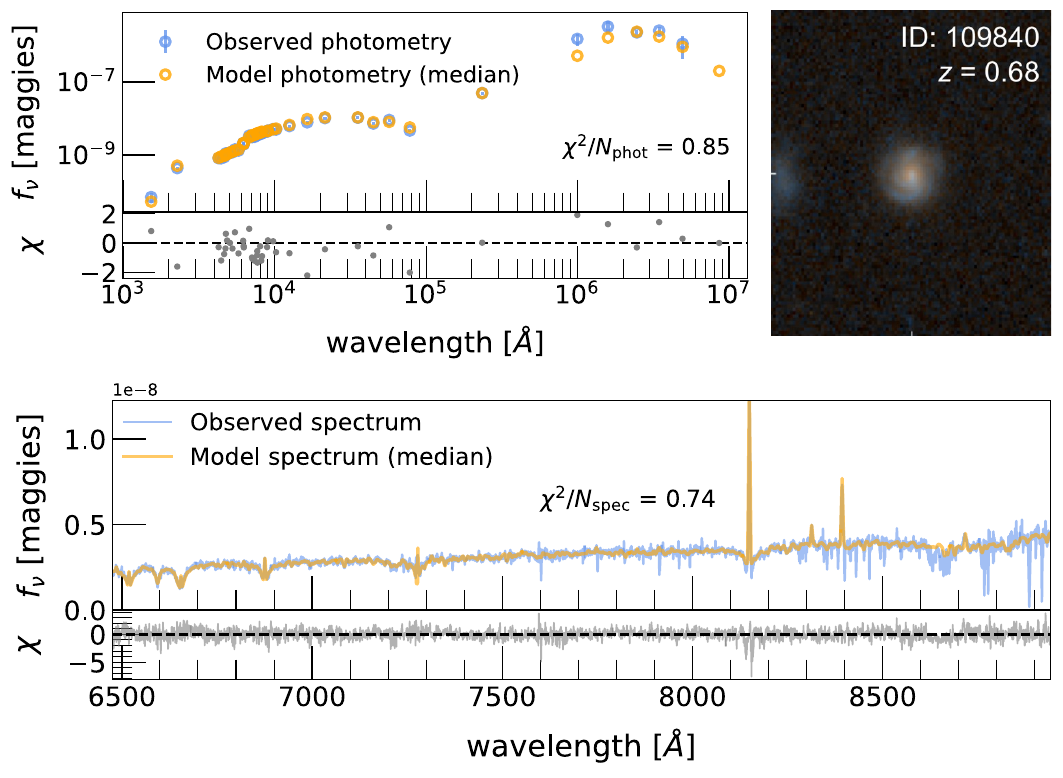}
    \caption{Observational data with the fitted {\typewriter Prospector} model for an example galaxy at $z = 0.67$. The top left panel shows the photometry; the top right panel shows the HST image; and the bottom panel shows the spectroscopic data (S/N $= 27.9 \Angstrom^{-1}$). The observed photometry and spectrum are plotted in blue, and the model photometry and spectrum are shown in orange. The associated $\chi$ values of the fit (defined as (model $-$ data)$/$observational uncertainty) are shown in the lower subplots of the top left and bottom panels. The overall reduced $\chi^2$ for the photometric and spectroscopic fits are 0.85 and 0.74, respectively.}
    \label{fig:example galaxy}
\end{figure*}

\begin{figure*}
    \centering
    \includegraphics[width=0.745\textwidth]{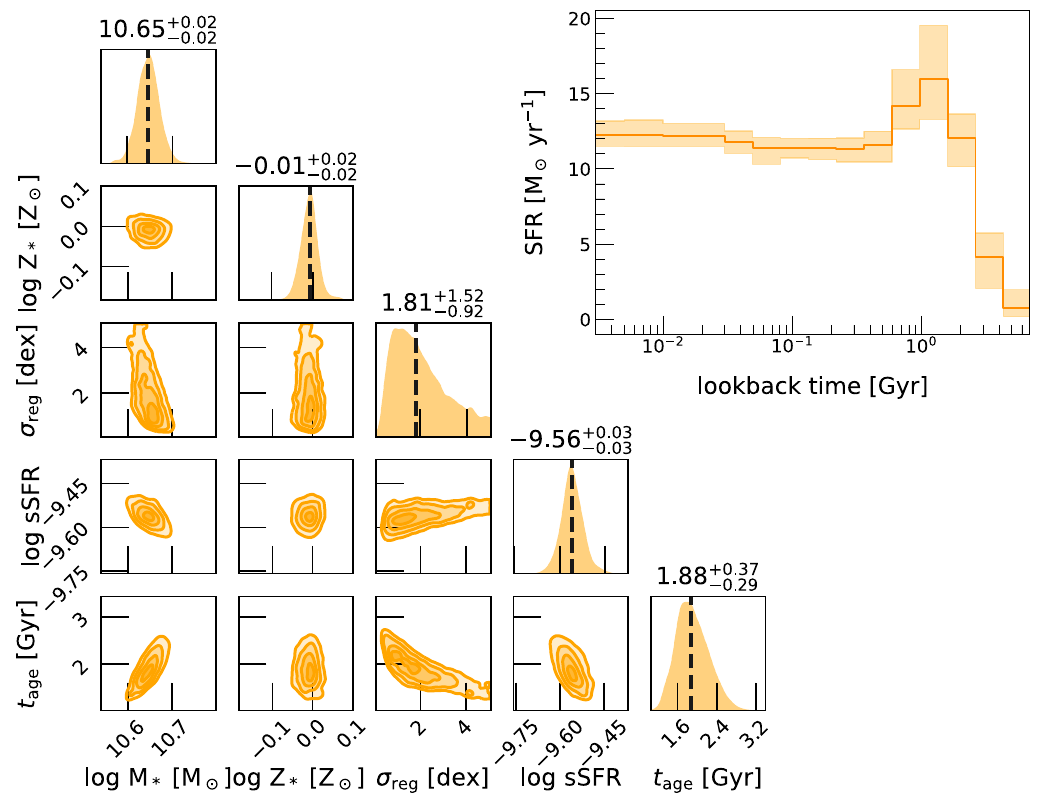}
    \caption{Joint posterior plot of key quantities and SFH for the galaxy shown in Figure \ref{fig:example galaxy}. The quantities in the corner plot are stellar mass ($\Mstar$), stellar metallicity ($\mathrm{Z_*}$), the PSD parameter $\sigreg$, specific star-formation rate (sSFR), and the mass-weighted stellar age ($\tage$). The posterior SFH is show in the upper right insert, where the solid blue line denotes the median and the shaded region covers the 16th$-$84th percentiles.}
    \label{fig:example fit}
\end{figure*}

We show the observational data for one example galaxy in our sample, along with the fitted model from {\typewriter Prospector}, in Figure \ref{fig:example galaxy}. This particular galaxy has a redshift of $z = 0.68$, and a median S/N of $27.9 \Angstrom^{-1}$ in the spectroscopic data. The model provides a good fits to the data --- the residuals are distributed around zero, and the reduced $\chi^2$ for the photometry and spectroscopy are 0.85 and 0.74, respectively.

The resulting posteriors for some key quantities of this fit are shown in Figure \ref{fig:example fit}. The parameters displayed in the joint posterior plot are the stellar mass ($\Mstar$), stellar metallicity ($\rm{Z}_*$), the PSD parameter $\sigreg$, specific star-formation rate (sSFR), and the mass-weighted stellar age ($\tage$). We find a stellar mass of $\log \Mstar/\rm{M}_\odot = 10.65 \pm 0.02$ and a stellar metallicity consistent with solar, $\rm{Z}_*/\rm{Z}_\odot = -0.01 \pm 0.02$. The variability in the gas inflow and cycling processes in this system is $\sigreg/\mathrm{dex} = 1.81 ^{+1.52}_{-0.92}$. The age of this system is $\tage/$Gyr $= 1.88^{+0.37}_{-0.29}$.  

Additionally, we classify the galaxies in this work into evolutionary regimes using the mass-doubling number \citep[following][]{Tacchella2022Halo7D}, which is defined as
\begin{equation}
\label{eq:doubling time eq}
    \mathcal{D}(z) = \mathrm{sSFR}(z) \times t_H(z),
\end{equation}
where sSFR$(z)$ and $t_H(z)$ are the specific star-formation rate of a galaxy and the age of the universe at the epoch of observation, respectively. $\mathcal{D}(z)$ represents, then, the number of times the stellar mass doubles within the age of the universe at redshift $z$, assuming the sSFR remains constant. We define a galaxy to be star-forming if $\mathcal{D}(z) > 1/3$, transitioning if $1/20 < \mathcal{D}(z) < 1/3$, and quiescent if $\mathcal{D}(z) < 1/20$. 

Looking at the example fit in Figure \ref{fig:example fit}, we find that this galaxy has a sSFR of log sSFR$/\mathrm{yr}^{-1} = -9.56 \pm 0.03$ and a doubling number of $\mathcal{D} = 2.05$, meaning it takes this system $\sim0.5\times$ the Hubble time (i.e., the age of the universe) to double its mass. Thus, it is classified as a star-forming galaxy.

\section{Reconstructed galaxy properties}
\label{sec:reconstructed galaxy properties}

Before diving into the nature of the SFMS, it is useful to first understand the basic properties of the galaxies in our sample, both as a sanity check on the SED-fitting procedure, as well as to establish the broader context of the star-forming galaxies we are ultimately interested in. Thus, in this section, we briefly examine our full set of inferred SFHs (Section \ref{sec:SFHs}), stellar ages (Section \ref{sec:mass-weighted ages}), and $\sigreg$ values (i.e., long-timescale SFH variability; Section \ref{sec:sigma_reg}), highlighting interesting details about the star-forming sample in particular.\footnote{All galaxy properties discussed hereafter (e.g., stellar mass, SFR) are estimated from the {\typewriter Prospector} fits described in this work, not to be confused with the nominal values in the LEGA-C DR3 catalog.} 

\subsection{Star-formation histories}
\label{sec:SFHs}

\begin{figure*}
    \centering
    \includegraphics[width=0.85\textwidth]{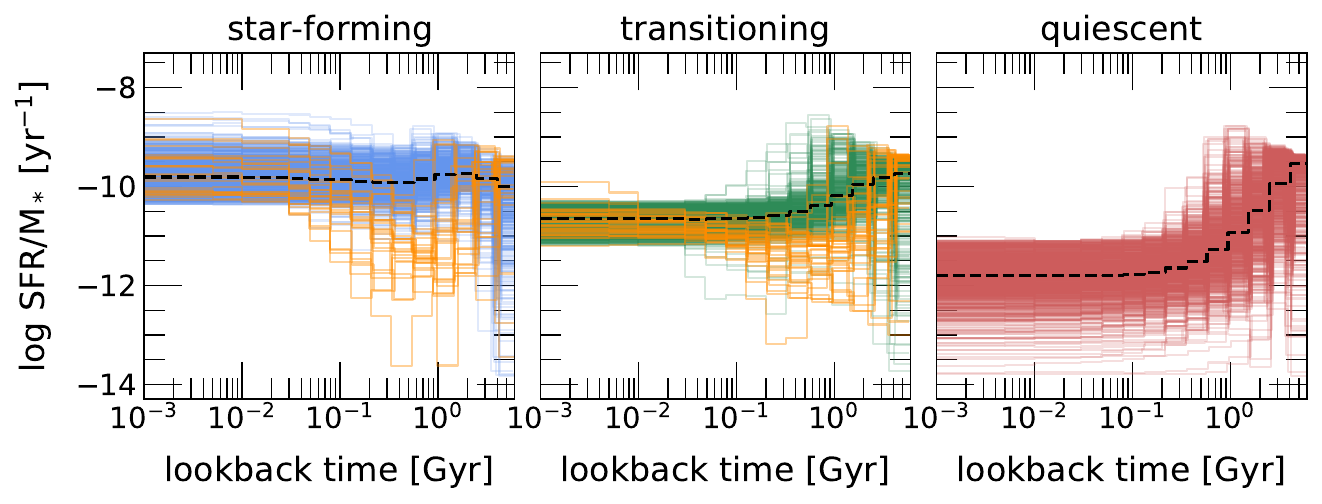}
    \caption{Reconstructed SFHs for all the galaxies in our sample, plotted as log SFR (normalized by final observed stellar mass) versus lookback time. The SFHs are separated according to evolutionary regime (as defined in Section \ref{sec:galaxy sample}, Eq. \ref{eq:doubling time eq}), with star-forming galaxies shown in blue in the leftmost panel, transitioning galaxies in green in the center panel, and quiescent galaxies in red in the rightmost panel. We show the median SFH in each category with a black dashed line. Furthermore, the galaxies that have been identified as rejuvenating are highlighted in orange. The median stellar mass in each of these categories is $\log \Mstar/\Msun = 10.99^{+0.25}_{-0.29}$ (star-forming), $\log \Mstar/\Msun = 11.22^{+0.24}_{-0.25}$ (transitioning), $\log \Mstar/\Msun = 11.30^{+0.20}_{-0.25}$ (quiescent), and $\log \Mstar/\Msun = 11.36^{+0.21}_{-0.36}$ (rejuvenating). In each evolutionary phase, individual galaxies' SFHs span a large range of tracks through the SFR-lookback time space; yet, there is still an overall coherence in the shapes of SFHs from regime to regime.}
    \label{fig:all_sfhs}
\end{figure*}

Figure \ref{fig:all_sfhs} presents the SFHs of all of the galaxies in our sample, plotted as log SFR (normalized by final observed stellar mass) versus lookback time and split by evolutionary regime --- star-forming (left-hand panel, shown in blue), transitioning (center panel, green), and quiescent (right-hand panel, red). The median SFH in each category is shown with a black dashed line. Galaxies are classified into these categories using the mass-doubling diagnostic specified by Equation \ref{eq:doubling time eq} in Section \ref{sec:fitting results}. We find that 647 galaxies in our sample are star-forming, 416 are transitioning, and 865 are quiescent. Furthermore, we identify 63 galaxies that have \textit{rejuvenated}, 25 of which are currently star-forming, and the remaining 38 transitioning. We consider a galaxy to be rejuvenating if it was quiescent in the past (i.e., at some point in its SFH, $\mathcal{D} < 1/20$; see Equation \ref{eq:doubling time eq} for a definition of $\mathcal{D}$) but has $\mathcal{D} > 1/20$ at the time of observation. The SFHs of these rejuvenating galaxies are shown in orange. Hereafter, ``star-forming'' and ``transitioning'' galaxies refer to star-forming and transitioning galaxies \textit{not including} rejuvenating systems.

There is a wealth of information condensed in, and analysis to be done on, this figure alone (e.g., on what timescales do galaxies quench? How long do they spend in the transition period?). However, investigating the diverse pathways through which galaxies go from star-forming to quenched is not the focus of the paper, so we will only discuss this plot briefly and leave an in-depth analysis for future work.

It is clear to see that, regardless of classification, galaxies follow a wide range of paths through SFR--time space. Some star-forming galaxies formed the bulk of their mass in a steady-state over long timescales; others formed a significant fraction of their stellar mass in one or two bursts, occasionally even transitioning off the SFMS before reigniting their star formation later in life. SFHs are similarly varied in both the transitioning and quiescent populations. 

However, there are also trends in the SFHs of each evolutionary phase. Looking at the median SFH in each phase (dashed black lines), we see that quiescent galaxies typically formed the earliest, with SFRs that peak at large lookback times before declining into a dormant state. Conversely, star-forming galaxies, on average, have sustained SFRs of $\gtrsim 10~\Msun~\mathrm{yr}^{-1}$ on timescales of $\sim1$~Gyr. Transitioning galaxies, as the name may suggest, lie somewhere in between, with SFHs that tend to peak at early times (although not as strongly as the quiescent population) and tail off towards the present-day. We continue this discussion in a more quantitative manner in the proceeding section.

\subsection{Mass-weighted ages}
\label{sec:mass-weighted ages}

\begin{figure*}
    \centering
    \includegraphics[width=0.96\textwidth]{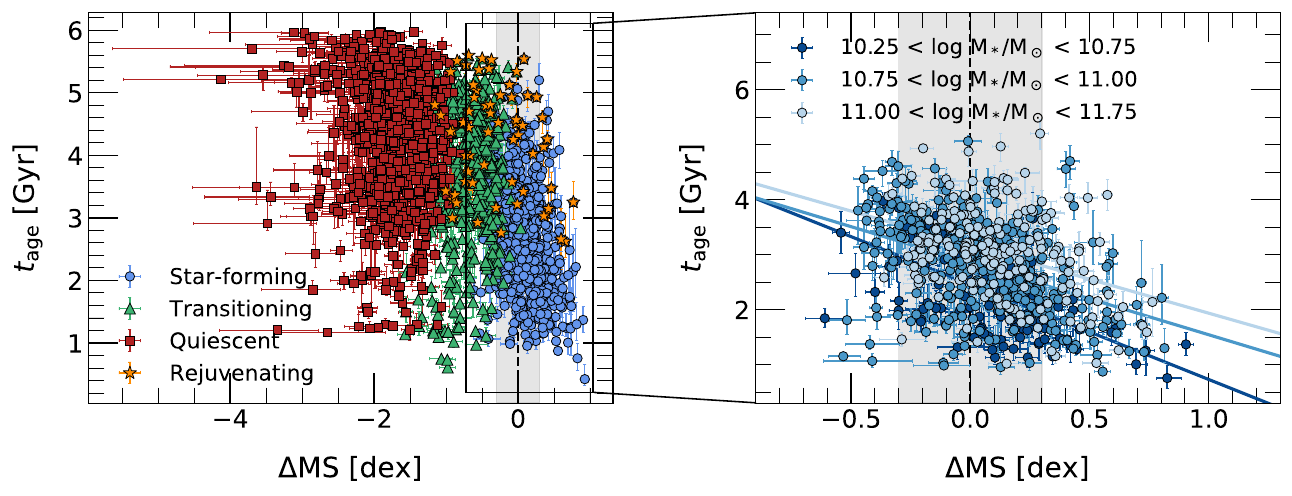}
    \caption{\textit{Left:} Mass-weighted age $t_{\rm{age}}$ as a function of distance from the SFMS (based on the SFR averaged over 30 Myr) for all galaxies in our sample. The blue circles, green triangles, red squares, and orange stars indicate star-forming, transitioning, quiescent, and rejuvenating galaxies, respectively. The vertical black line marks the SFMS ($\Delta \rm{MS} = 0$), and the gray band shows the scatter of the SFMS (roughly 0.3 dex). There is a weak age gradient across the SFMS, hinting at a long-term correlation in galaxies' position on the main sequence. \textit{Right:} A zoom-in on the star-forming galaxies in the left-hand panel. We bin the star-forming galaxies by stellar mass, with dark blue corresponding to galaxies in the lowest mass bin ($10.25 < \log \Mstar/\Msun < 10.75$), medium blue to the middle mass bin ($10.75 < \log \Mstar/\Msun < 11.0$), and light blue to the highest mass bin ($11.0 < \log \Mstar/\Msun < 11.75$). The solid lines indicate the best-fit of the $t_{\rm{age}} - \DeltaMS$ relationship in each mass bin. The age gradient remains present at fixed stellar mass. Additionally, massive systems are typically older than less massive systems.}
    \label{fig:mwa}
\end{figure*}

We can use mass-weighted age ($\tage$), which is directly computed from our SFHs, to distill the overall coherence in the shapes of star-forming, transitioning, and quiescent SFHs into a simple parameter. Figure \ref{fig:mwa} shows $\tage$ as a function of distance from the SFMS in logarithmic space ($\Delta \rm{MS}$) for all (left) and star-forming (right) galaxies in our sample, where $\Delta \rm{MS}$ is defined as
\begin{equation}
\label{eq:delta MS}
     \Delta \mathrm{MS} \equiv \log \big{(} \mathrm{SFR_{30}}/\mathrm{SFR_{MS}}(\Mstar, z) \big{)}\,
\end{equation}
for a galaxy of stellar mass $\Mstar$ at a redshift of $z$. Here, SFR$_{30}$ is calculated from the SFR averaged over 30 Myr and SFR$_{\rm{MS}}$ uses the SFMS definition of \citet{Leja2022}. A vertical dashed black line marks the SFMS, and the gray band indicates the $\sim 0.3$ dex scatter. 

Let us first focus on the galaxy sample at large (left panel of Figure \ref{fig:mwa}). Here, star-forming, transitioning, quiescent, and rejuvenating systems are distinguished by blue circles, green triangles, red squares, and orange stars, respectively. Although galaxies in every evolutionary regime span a range of ages, on average, quiescent systems are the oldest, with $\tage = 4.46^{+0.91}_{-1.20}$ Gyr, star-forming systems are the youngest, with $\tage = 2.85^{+0.76}_{-1.04}$ Gyr, and transitioning galaxies sit in between, with $\tage = 3.79^{+0.91}_{-1.44}$ Gyr. This is consistent with both our reconstructed median SFHs from the previous section, as well as the expectation that quiescent and transitioning galaxies have been in, or are moving into, an extended epoch of low star formation, while star-forming galaxies are still actively building up their stellar masses.\footnote{It is also worthwhile to remember that the galaxies in this sample span $0.6 \leq z \leq 1$, so redshift--age trends are also buried in Figure \ref{fig:mwa}. The age difference between evolutionary regimes becomes more pronounced once that is taken into account.}

Additionally, the 63 rejuvenating galaxies identified by our (rather strict) rejuvenation criteria are fairly old given their positions relative to the main sequence. Despite being at roughly the same distance from the SFMS as the star-forming and transitioning populations (note the $\Delta \rm{MS}$ values of the orange stars in Figure \ref{fig:mwa}), the rejuvenating systems have an average mass-weighted age of $\tage = 4.34^{+0.78}_{-1.29}$ Gyr, on par with the age of the quiescent population. This is consistent with the rejuvenating galaxies having formed most of their stars in an early burst (as seen in their SFHs in Figure \ref{fig:all_sfhs}) before climbing back up onto the SFMS. Together, this implies that 1) rejuvenation is uncommon, given that only 3.3\% of our galaxy sample has experienced a rejuvenation event, and 2) the amount of mass formed in such events is not significant ($\lesssim 10\%$). Both the rarity of rejuvenating galaxies and the insignificance of rejuvenation events to the stellar mass build-up of these systems is consistent with previous observational measurements \citep{Chauke2019, Tacchella2022Halo7D, Tanaka2023}.

Now we can hone in on the star-forming systems (right panel of Figure \ref{fig:mwa}). We split the star-forming sample into three stellar mass bins --- $10.25 < \log \Mstar/\Msun < 10.75$, $10.75 < \log \Mstar/\Msun < 11.0$, and $11.0 < \log \Mstar/\Msun < 11.75$ --- which color-code the galaxies by increasingly lighter shades of blue. The solid blue lines indicate the best-fit of the $t_{\rm{age}} - \DeltaMS$ relationship in each mass bin. First of all, the most massive galaxies on the SFMS also tend to be the oldest, while the least massive galaxies are on average the youngest. The median mass-weighted age in the $10.25 < \log \Mstar/\Msun < 10.75$ mass bin is $2.22^{+1.01}_{-0.73}$ Gyr; in the $10.75 < \log \Mstar/\Msun < 11.0$ mass bin, it is $2.83^{+0.80}_{-0.99}$ Gyr; and in the $11.0 < \log \Mstar/\Msun < 11.75$ mass bin, the median age is $3.08^{+0.71}_{-0.88}$ Gyr. This makes sense --- in order to achieve very high stellar masses, galaxies likely needed to maintain high SFRs over extended periods of time, increasing the mass-weighted age, whereas less massive systems could have formed significant fractions of their masses in a recent burst.

Secondly, there is a weak age gradient across the main sequence ridgeline. Galaxies above the SFMS tend to be younger ($\tage = 2.34^{+0.94}_{-0.71}$ Gyr) than those below it ($\tage = 3.29^{+0.61}_{-0.92}$ Gyr). This trend remains when looking at a fixed stellar mass bin. The galaxies furthest above the SFMS are not only the youngest, but also typically less massive, and can be explained by recent starbursts. We find that this age gradient is strongest in the lowest-mass bin (where the Pearson correlation coefficient between $\tage$ and $\DeltaMS$ is $R = -0.59$) and is somewhat weaker in the intermediate and highest-mass bins ($R = -0.42$ and $-0.36$, respectively). This is also evident from the slopes of the best-fit lines in Figure \ref{fig:mwa}, which become less steep with increasing mass bin.

These two pieces of information foreshadow two interesting directions of investigation: 1) the overall trend in age across the SFMS hints at a long-term correlation as to whether galaxies are above or below the SFMS ridgeline, and 2) the extreme starbursts present in the low-mass bin lend additional credence to the idea that lower-mass galaxies experience larger variations in their SFHs than high-mass galaxies \citep[e.g.,][]{Weisz2012, Guo2016}. Both of these ideas will be discussed thoroughly in Section \ref{sec:characterizing the SFMS}.

\subsection{$\sigreg$ --- Long-term SFH variability}
\label{sec:sigma_reg}

\begin{figure}
    \centering
    \includegraphics[width=0.48\textwidth]{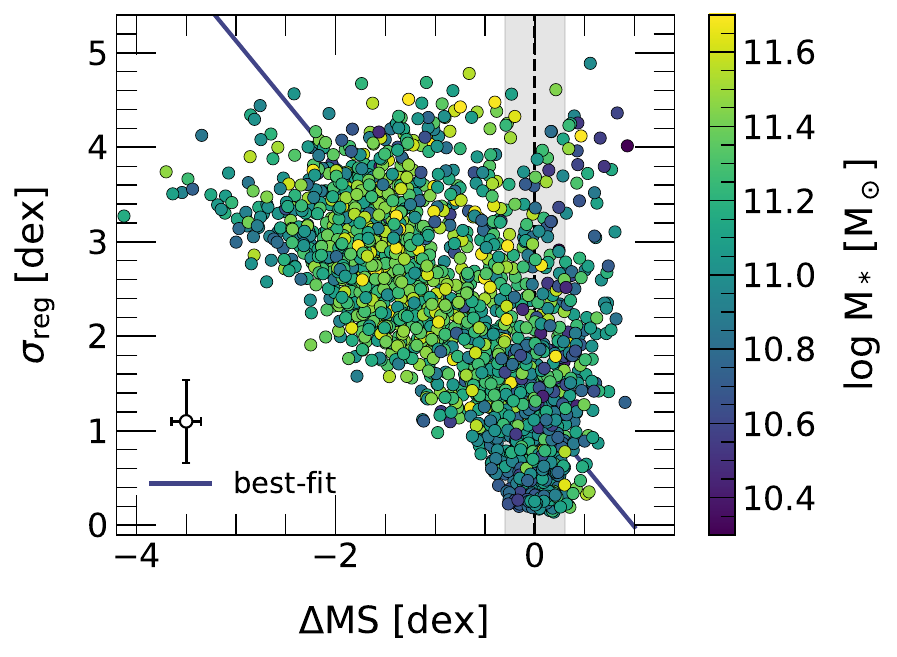}
    \caption{$\sigreg$ versus $\DeltaMS$ for all the galaxies in our sample, color-coded by stellar mass (log M$_*$). The characteristic error bar is shown in the lower left. The solid indigo line marks the best-fit between $\sigreg$ and $\DeltaMS$ (accounting for uncertainties in both variables). The black dashed line marks the SFMS ($\DeltaMS = 0$), and the gray band shows its $\sim0.3$ dex scatter. $\sigreg$ tends to increase as galaxies move away from the SFMS ridgeline, with quiescent galaxies displaying the largest values, implying that long-timescale SFH fluctuations are more dominant in quiescent galaxies than star-forming systems.}
    \label{fig:sfr-mass sigreg}
\end{figure}

Because the SFHs of our galaxies were fit using the stochastic SFH prior \citep{Wan2024}, we are actually able to constrain the power contained in long-timescale SFH fluctuations for each galaxy. This is quantified by the parameter $\sigreg$ in our SED-fitting model. Figure \ref{fig:sfr-mass sigreg} plots $\sigreg$ as a function of distance from the SFMS ($\DeltaMS$) , color-coded by stellar mass (log M$_*$). The best-fit linear relationship between $\sigreg$ and $\DeltaMS$ is shown with a solid indigo line.

There is a clear trend in $\sigreg$ across the sample --- galaxies on the SFMS tend to have low values of $\sigreg$, and moving away from the SFMS, $\sigreg$ increases, with the largest values seen in quiescent galaxies (the Pearson correlation coefficient between $\sigreg$ and $\DeltaMS$ is $R = -0.60$). On average, we find $\sigreg = 1.24^{+1.40}_{-0.82}$ dex in star-forming galaxies, $\sigreg = 2.08^{+1.38}_{-0.85}$ dex in transitioning galaxies, and $\sigreg = 2.93^{+1.04}_{-0.86}$ dex in quiescent galaxies. This indicates that the amount of power contained in long-timescale ($\gtrsim 1$ Gyr) SFH fluctuations is greater in quiescent galaxies than in star-forming ones. This makes sense given that essentially all quiescent galaxies have experienced a many orders-of-magnitude reduction in their SFRs over their lifetimes, and is furthermore consistent with previous findings from simulations \citep{Iyer2020}. 

Additionally, there is significant scatter towards larger values of $\sigreg$ in the star-forming galaxies that are further away from the SFMS (larger values of $|\DeltaMS|$). Galaxies more than 0.3 dex above the SFMS have an average $\sigreg = 1.83^{+1.60}_{-0.95}$ dex, while those within the 0.3 dex envelope of the SFMS have $\sigreg = 1.06^{+1.27}_{-0.73}$ dex. This provides evidence that galaxies significantly above the SFMS ridgeline are experiencing burstier star-formation than those near the SFMS neighborhood.

Unfortunately, $\sigreg$ is not the most useful metric for investigating movement about the SFMS. As seen in \citet{Wan2024}, the current data and state of SED-modeling in {\typewriter Prospector} do not allow for robust constraints to be placed on any of the PSD parameters outside of $\sigreg$. This manifestly means that we cannot estimate the PSD which underlies our galaxy sample's SFHs from SED-fitting. 

Another implication of this is that $\sigreg$ effectively sets the width of the SFH prior and dictates what range of SFRs a galaxy can undergo. Indeed, there is a triangle of parameter space of small $\sigreg$ and negative $\DeltaMS$ that seems to be disallowed by the data --- if a galaxy quenches, its SFH model must necessarily have a large value of $\sigreg$ in order to fit the data. In other words, $\sigreg$ compresses \textit{all} the variability information across a galaxy's entire lifetime into a single value, such that it becomes difficult to distinguish whether a $\sigreg$ is large due to a short-timescale starburst or a large-amplitude but gradual change in SFR, despite those two scenarios arising from very different physical processes.

But fret not! There is hope yet. If we believe that the SFHs of our star-forming galaxies are well-constrained (i.e., both accurate and precise), which we generally do (see Section 4 of \citealt{Wan2024} for the relevant recovery tests), then it is possible to construct other metrics to address questions relating to the fluctuations in galaxy SFHs, timescales, and the nature of the SFMS.

\subsection{Other galaxy parameters}
\label{sec:other params}

There are a large number of galaxy properties (see Table \ref{tab:priors}) inferred for the systems in this work that we do not mention here. While we defer the detailed analysis of, e.g., dust parameters and metallicities to future work, it is important to acknowledge that reliably estimating these parameters is critical to reliably estimating galaxy SFHs. The recovery tests performed in \citet{Wan2024} have shown that the {\typewriter Prospector} model used in this analysis is able to accurately recover key stellar population parameters, including dust attenuation and stellar metallicity, when applied to LEGA-C-resolution spectrophotometric data.

We verify there are no strong covariances between dust, age, and metallicity in the resulting posterior distributions for our galaxies. This degeneracy, typically a concern for any SED-fitting analysis, is able to be broken with the joint fitting of high-resolution LEGA-C optical-wavelength spectra and photometry from the COSMOS2020 catalog, which spans the (rest) UV to near-IR (see Section \ref{sec:data} for details). The LEGA-C spectra, although narrower in wavelength range compared to the photometry, give us access to key features like the Balmer lines from H$\beta$ (at $4861~\Angstrom$) to H$\delta$ (at $4102~\Angstrom$), the Calcium H and K absorption lines (at $3934~\Angstrom$ and $4455~\Angstrom$, respectively), the CN line (at $4160~\Angstrom$), and several other Mg, Ca, and Fe lines \citep{LEGAC2021}. Thus, the broad wavelength coverage of the photometry gives us a strong lever arm on dust attenuation and stellar mass, while the age and metallicity of a galaxy are mainly pinned down by the spectroscopy (see Appendix B of \citealt{Tacchella2022Halo7D} for a more thorough discussion on this).

For completeness, we provide a comparison between the galaxy stellar masses estimated in this work and the stellar masses from LEGA-C DR3 catalog and \citet{Cappellari2023} in Appendix \ref{app:comparison}, as well as an overview of our recovered dust attenuation and stellar metallicity measurements in Appendix \ref{app:dust and met}. Additionally, results of the entire suite of parameters fit by {\typewriter Prospector} for the galaxies in this sample are available as supplementary material.

\section{Characterizing the star-forming main sequence}
\label{sec:characterizing the SFMS}

As advertised, the primary motivation for this work is to build a better understanding of the nature of the star-forming main sequence. Do galaxies evolve \textit{along} the SFMS, or do they evolve \textit{across} it? These two scenarios imply fundamentally different interpretations of the main sequence. In the former case, similar-mass galaxies ``grow up'' together, i.e., the SFMS is an evolutionary sequence \citep{Peng2010, Behroozi2013}; in the latter case, the SFMS is not an attractor solution, but rather an observational coincidence \citep{Gladders2013, Abramson2016}.

An unfortunate fact of life for astronomers is that the timescales on which humans operate are much shorter than the timescales governing galaxies. As such, it is impossible to conduct a longitudinal galaxy evolution study (i.e., to measure how the properties of the same set of galaxies change over time). However, it is possible to infer the historical properties (e.g., SFHs) of the galaxies we observe in our thin snapshot of the universe and construct an equivalent formulation of the SFMS problem through the lens of SFH variability. Namely, if the SFMS is an evolutionary track, then the internal processes at work in galaxies, such as the creation and destruction of GMCs, serve as modulators of star-formation activity. This would mean that systems on the main sequence experience primarily short-term ($\sim$tens to hundreds of Myr) fluctuations in their SFHs, and the scatter in the main sequence relation encodes these short-timescale variations. On the other hand, if the SFMS is just observational happenstance, then we would expect galaxy SFHs to primarily vary on long ($\gtrsim 1$ Gyr) timescales and the SFMS scatter to reflect this long-term diversity.

Thus, we approach the task of distinguishing between the ``evolutionary'' and ``coincidental'' interpretations of the SFMS in two distinct, but complimentary, ways. First, we measure the SFR dispersion in our sample of star-forming galaxies over a range of timescales (Section \ref{sec:what's in a scatter}). Then, we decompose our galaxies' offsets from the SFMS into short-timescale and long-timescale components and compare their relative amplitudes (Section \ref{sec:what's in an offset}). 

\subsection{Scatter of the SFMS}
\label{sec:what's in a scatter}

It is clear that observed scatter in the SFMS ($\sigmaMS$) is a reflection of the amplitude of SFR variations that a galaxy population experiences. (If star-forming galaxies are able to span a large range of SFRs at a given epoch, then the measured $\sigmaMS$ will similarly be large.) A slightly more subtle point is that $\sigmaMS$ also depends on the timescale over which galaxy SFRs are evaluated. Because in a distant galaxy we cannot literally count the number of new stars being born at any given moment, any SFR we measure is necessarily an \textit{effective} SFR averaged over some timescale. The way in which $\sigmaMS$ behaves as a function of this averaging timescale carries information about how rapidly galaxies are changing their SFRs. If galaxy SFHs decorrelate very quickly (i.e., their SFRs vary on short timescales), then there will be a significant difference between $\sigmaMS$ measured over short timescales versus long timescales, since averaging SFR over longer periods will wash out the effects of short-term bursts, thereby decreasing the dispersion. If galaxy SFHs are generally steady, only experiencing fluctuations on extremely long timescales, then $\sigmaMS$ should be nearly independent of averaging timescale. 

To investigate this, we calculate $\sigmaMS$ with different SFR averaging timescales for our star-forming galaxy sample. We break the sample into three stellar mass bins: $10.25 < \log \Mstar/\Msun < 10.75$, $10.75 < \log \Mstar/\Msun < 11.0$, $11.0 < \log \Mstar/\Msun < 11.75$. In a given mass bin, we measure the SFRs of the galaxies over timescales of (0.01, 0.03, 0.1, 0.25, 0.5, 1.0, and 2.0) Gyr. To be explicit, we use the reconstructed SFH of each galaxy to calculate the total stellar mass formed over each of the desired timescales for a given galaxy. The total mass formed divided by its associated timescale results in the galaxy's SFR averaged over that timescale. $\sigmaMS$ at a given timescale is then defined as the standard deviation in the log SFRs measured over that averaging timescale for all the galaxies in the mass bin.

\begin{figure}
    \centering
    \includegraphics[width=0.48\textwidth]{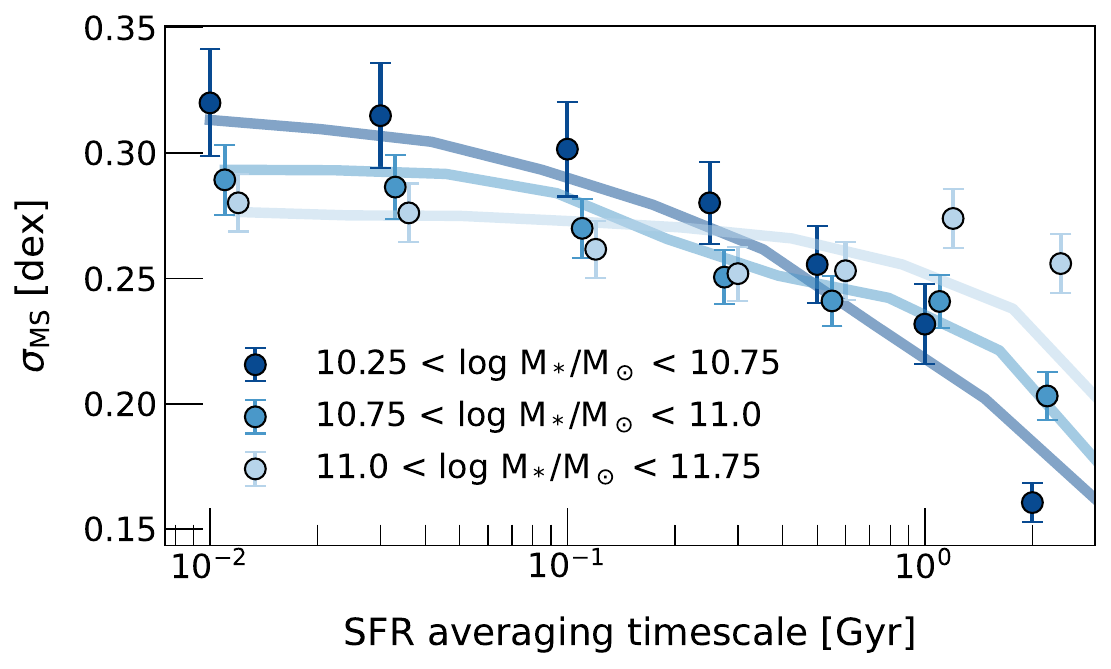}
    \caption{Scatter of the log SFRs of star-forming galaxies ($\sigmaMS$) in three bins of stellar mass, plotted as a function of timescale over which the SFRs were averaged. The dark blue points correspond to galaxies in the lowest mass bin ($10.25 < \log \Mstar/\Msun < 10.75$), medium blue to the middle mass bin ($10.75 < \log \Mstar/\Msun < 11.0$), and light blue to the highest mass bin ($11.0 < \log \Mstar/\Msun < 11.75$). Points are offset slightly on the $x$-axis for clarity. $\sigmaMS$ is largest in lower-mass galaxies when the SFR averaging timescale is short and declines rapidly with longer averaging timescales. In the highest-mass bin, $\sigmaMS$ shows a much weaker decrease with averaging timescale. Furthermore, scatter of the main sequence as a function of SFR averaging timescale can be used to constrain the underlying PSD \citep{TFC2020}. In solid lines, we show the analytically-derived relations between $\sigmaMS$ and averaging timescale as drawn from three different of PSDs. The line color indicates which mass bin is most closely matched. We find that $\sigdyn$ (i.e., the intrinsic stochasticity associated with short-timescale processes like the creation and destruction of GMCs) decreases with stellar mass, while $\sigreg$ (i.e., the intrinsic variability in the longer-timescale behavior of a galaxy's gas reservoir) increases. Additionally, $\taueq$, the correlation timescale of equilibrium gas cycling, increases with stellar mass.}
    \label{fig:ms scatter timescales}
\end{figure}

In Figure \ref{fig:ms scatter timescales}, we plot $\sigmaMS$ as a function of SFR averaging timescale. The dark blue points correspond to galaxies in the lowest mass bin, medium blue to the middle mass bin, and light blue to the highest mass bin. The mean observational uncertainty in log SFR, as a function of stellar mass and averaging timescale, has been deconvolved (i.e., subtracted in quadrature) from each $\sigmaMS$ measurement. We verify that in all instances, this is a small ($\lesssim 3$\%) effect.

Because all of the galaxies in this sample are massive ($\gtrsim 10^{10}~\Msun$) and around the same redshift ($z \sim 0.8$), we do not expect the normalization of $\sigmaMS$ versus timescale to be significantly different between stellar mass bins. Indeed, this is what we see --- the $\sigmaMS$ measured over $10 - 30$ Myr is $\sim0.3$ dex in all three mass bins, which is consistent with previous studies of the SFMS at $z \sim 0.8$ \citep[e.g.,][]{Noeske2007, Salim2007, Zahid2012}. There is a mild decrease in the scatter with stellar mass, from $\sim0.32$ dex in the $10.25 < \log \Mstar/\Msun < 10.75$ mass bin to $\sim0.28$ dex in the $11.0 < \log \Mstar/\Msun < 11.75$ mass bin, which indicates that lower-mass galaxies experience slightly larger SFR variations than higher-mass systems. However, this difference is not substantial.

There is, however, a substantial difference in the behavior of $\sigmaMS$ versus SFR averaging timescale between mass bins at long timescales. Where $\sigmaMS$ undergoes a clear decline with averaging timescale in the lowest-mass bin, $\sigmaMS$ remains nearly flat in the highest-mass bin. This implies that the timescales over which galaxy SFHs remain correlated increase with stellar mass. So if we consider the picture where star-forming galaxies fluctuate around the SFMS, this would mean that galaxies' SFRs typically remain within a $\pm 0.3$ dex envelope, with lower-mass galaxies oscillating within this envelope on shorter timescales than higher-mass galaxies.

\citet{TFC2020} showed that the scatter of the main sequence as a function of SFR averaging timescale can be used to constrain the underlying PSD (see Section 4.2.1 of the aforementioned paper). The intrinsic variability in galaxy SFHs, as well as the timescales over which the SFHs are correlated, affect both the normalization of the $\sigmaMS$--timescale relationship, as well as the steepness of the drop-off at large timescales. (The larger the amplitude of variation is, the larger $\sigmaMS$ will be. The longer the SFH correlation timescales are, the weaker the decline becomes.) While we leave making explicit constraints on the PSD for a future work, we present a preliminary investigation here.

To explore the space of PSDs that could give rise to the $\sigmaMS$ versus SFR averaging timescale trends that we observe in each mass bin, we sample SFHs from different PSDs over a grid of $\sigreg$, $\taueq$, $\tauin$, $\sigdyn$, and $\taudyn$ using the Gaussian process implementation of \citet{Iyer2024}. We then compute the scatter in log SFR obtained from these analytical SFHs, averaging the SFRs over timescales ranging from $0.01-2$ Gyr. We show the case which most closely matches the observed relation, as determined by a simple chi-squared minimization calculation, in each mass bin. In the lowest-mass bin, the set of PSD parameters which most resembles the observed $\sigmaMS$ versus averaging timescale trend is $(\sigreg, \taueq, \tauin, \sigdyn, \taudyn) =$ (0.13 dex, 0.25 Gyr, 0.5 Gyr, 0.35 dex, 0.1 Gyr); in the intermediate bin, the PSD parameters are $(\sigreg, \taueq, \tauin, \sigdyn, \taudyn) =$ (0.28 dex, 0.5 Gyr, 1.5 Gyr, 0.25 dex, 0.1 Gyr); and lastly, in the highest-mass bin, the PSD parameters are $(\sigreg, \taueq, \tauin, \sigdyn, \taudyn) =$ (0.62 dex, 2.0 Gyr, 7.0 Gyr, 0.15 dex, 2.0 Gyr). We find that $\sigdyn$ (i.e., the intrinsic stochasticity associated with short-timescale processes like the creation and destruction of GMCs) decreases with stellar mass, while $\sigreg$ (i.e., the intrinsic variability in the longer-timescale behavior of a galaxy's gas reservoir) increases. Additionally, $\taueq$ and $\tauin$, the correlation timescales of equilibrium gas cycling and gas inflow, increase with stellar mass. This provides evidence that lower-mass galaxies do, in fact, experience burstier star formation than more massive galaxies. Additionally, it means that galaxies whose SFHs are correlated over short and long timescales are both represented on the SFMS.

\subsection{Offsets from the SFMS}
\label{sec:what's in an offset}

To further investigate the relative contributions of long- and short-term SFR fluctuations to the overall SFH trajectories of our star-forming galaxies, we turn to their offsets from the SFMS ($\DeltaMS$). If a galaxy is observed to be at a position $\DeltaMS$ from the SFMS ridgeline (defined in Equation \ref{eq:delta MS}), there are two generic possibilities for how it got there. It could be passing through the point $\DeltaMS$ as it oscillates back and forth along the SFMS, or it could be intersecting the point as it passes through the SFMS on a one-way journey. Therefore, it makes sense to think about a galaxy's current position relative to the SFMS as a combination of a long-timescale offset $\DeltaL$ and a short-timescale offset $\DeltaS$ such that $\DeltaMS = \DeltaS + \DeltaL$. 

\begin{figure*}
    \centering
    \includegraphics[width=0.8\textwidth]{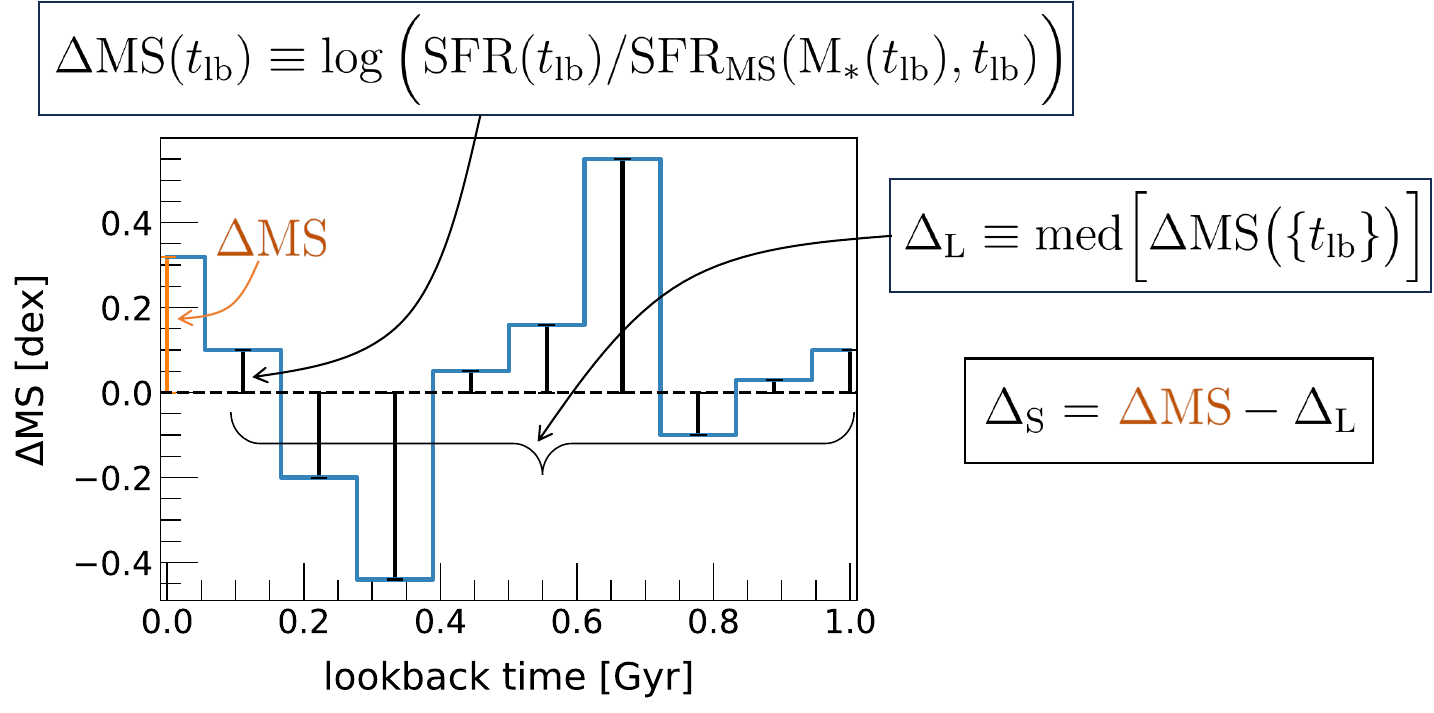}
    \caption{A cartoon depiction of how $\Delta_{\rm{L}}$ and $\Delta_{\rm{S}}$ are calculated in our $\DeltaMS$ decomposition analysis. We assume that $\DeltaMS$, i.e. a galaxy's \textit{current} position relative to the SFMS in logarithmic space (defined in Equation \ref{eq:delta MS}), arises from a combination of a long-timescale offset $\DeltaL$ and a short-timescale offset $\DeltaS$ such that $\DeltaMS = \DeltaS + \DeltaL$. $\DeltaL$ is defined to be the median \textit{historical} offset from the SFMS ($\DeltaMS(t_{\rm{lb}})$; defined in Equation \ref{eq:historical deltaMS}) measured at ten evenly spaced points over a galaxy’s SFH from a lookback time of 0.1 to 1 Gyr; and $\DeltaS$ is the difference between a galaxy's \textit{current} $\DeltaMS$ and $\DeltaL$. The quantity $|\DeltaS| - |\DeltaL|$ then quantifies the relative amplitudes of short- and long-term fluctuations in a galaxy's SFH. If $|\DeltaS| - |\DeltaL| > 0$, that means short-timescale variations play a more important role than long-timescales ones in determining the galaxy's current position on the SFMS. Conversely, if $|\DeltaS| - |\DeltaL| < 0$, then short-timescale variations are subdominant to long-timescale variations. In the cartoon SFH shown in this figure, $\DeltaMS = 0.32$ dex and $\DeltaL = 0.075$ dex; thus, $\DeltaS = 0.245$ dex. This means that $|\DeltaS| - |\DeltaL| = 0.17$ dex, i.e. this galaxy's position on the SFMS is primarily due to short-timescale variations in its SFH.}
    \label{fig:timescales_cartoon}
\end{figure*}

We define $\DeltaL$ to be the median \textit{historical} offset from the SFMS measured at ten evenly spaced points over a galaxy’s SFH, from a lookback time of 0.1 to 1 Gyr. In other words, $\DeltaL \equiv \mathrm{med} \big[ \DeltaMS(\{t_{\rm{lb}}\}) \big]$, where 
\begin{equation}
    \label{eq:historical deltaMS}
    \DeltaMS(t_{\mathrm{lb}}) \equiv \log \Big( \mathrm{SFR}(t_{\mathrm{lb}}) / \mathrm{SFR_{MS}}\big( \mathrm{M}_*(t_{\mathrm{lb}}), t_{\mathrm{lb}} \big) \Big).
\end{equation}
If a galaxy oscillates around the SFMS on short ($<$ 1 Gyr) timescales, then $\DeltaL$ should be close to zero; otherwise, $\DeltaL$ will have a nonzero value. $\DeltaS$ is then just the difference between a galaxy's \textit{current} $\DeltaMS$ and $\DeltaL$, its \textit{median historical} $\DeltaMS$. In this way, $\DeltaS$ captures any offset from the SFMS that cannot be explained by a long-timescale effect. Figure \ref{fig:timescales_cartoon} provides a cartoon depiction of how $\DeltaL$ and $\DeltaS$ are calculated. In the example SFH shown in this figure, $\DeltaMS = 0.32$ dex and $\DeltaL = 0.075$ dex; thus, $\DeltaS = 0.245$ dex. This means that $|\DeltaS| - |\DeltaL| = 0.17$ dex, i.e., this galaxy's position on the SFMS is primarily due to short-timescale variations in its SFH.

Importantly, $\DeltaL$ and $\DeltaS$ are constructed in such a way that the metric $|\DeltaS| - |\DeltaL|$ quantifies the relative amplitudes of short- and long-term fluctuations in a galaxy's SFH. If $|\DeltaS| - |\DeltaL| > 0$, then that means short-timescale SFR variations play a more important role than long-timescales ones in determining the galaxy's current position on the SFMS. Conversely, if $|\DeltaS| - |\DeltaL| < 0$, then short-timescale SFR fluctuations are subdominant to long-timescale ones.

\begin{figure*}
    \centering
    \includegraphics[width=0.95\textwidth]{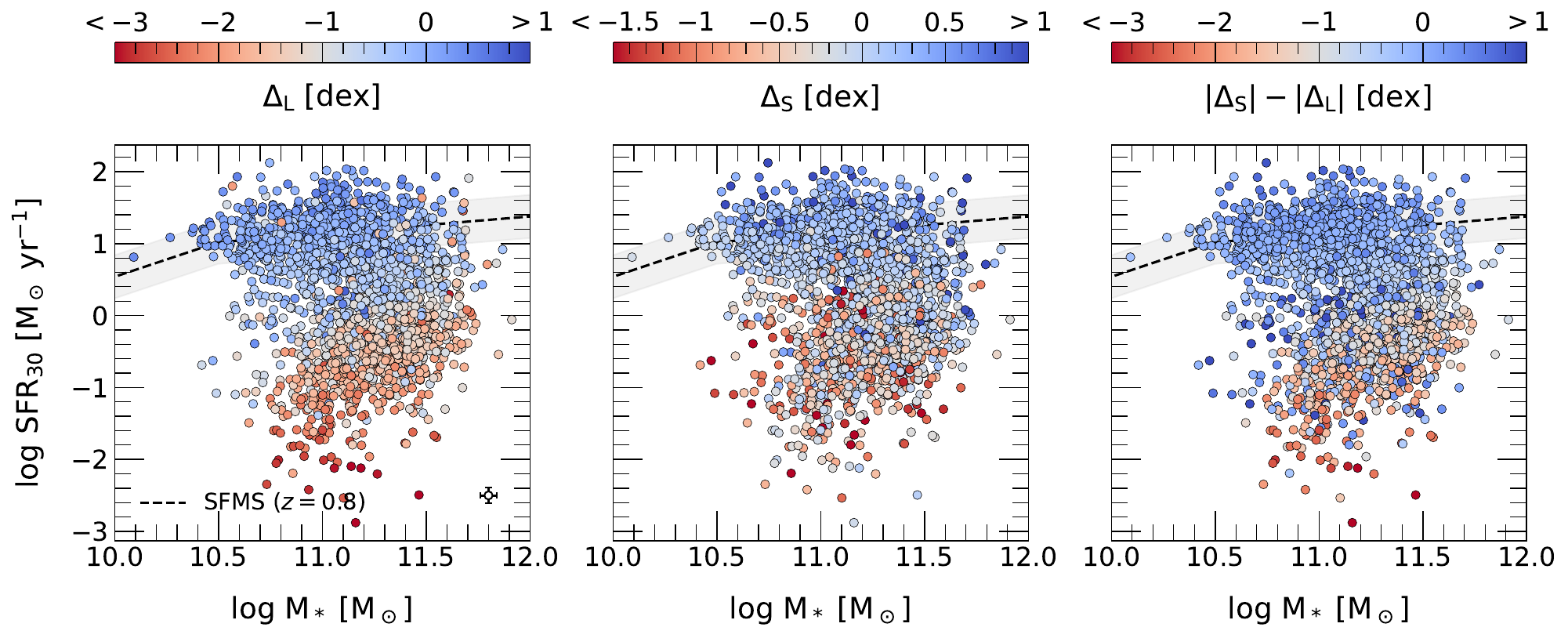}
    \caption{All galaxies in our sample plotted on the $\rm{SFR}-\Mstar$ plane, color-coded by (from left to right) $\Delta_{\rm{L}}$, $\Delta_{\rm{S}}$, and $|\Delta_{\rm{S}}| - |\Delta_{\rm{L}}|$. The star-forming main sequence from \citet{Leja2022} is marked with a black dashed line, and its scatter (roughly 0.3 dex) is shaded in gray. $\DeltaL$ represents a galaxy's median offset from the SFMS over the last 1 Gyr, and $\DeltaS$ represents a galaxy's offset from the SFMS on timescales $< 30$~Myr. $|\Delta_{\rm{S}}| - |\Delta_{\rm{L}}|$ compares the amplitude of these short- and long-term deviations from the SFMS. Long-timescale SFH variations are more important in determining the positions of quiescent galaxies relative to the SFMS (i.e., $|\Delta_{\rm{S}}| - |\Delta_{\rm{L}}| < 0$ in the quiescent population), while both short- and long-timescale variations contribute to the star-forming population ($|\Delta_{\rm{S}}| - |\Delta_{\rm{L}}| \sim 0$) on the main sequence.}
    \label{fig:timescales}
\end{figure*}

We can verify that this is true using Figure \ref{fig:timescales}, which shows the galaxies in our sample on the SFR$-\Mstar$ plane, color coded by (from left to right) $\DeltaL$, $\DeltaS$, and $|\DeltaS| - |\DeltaL|$. The \citet{Leja2022} parametrization of the SFMS is overplotted with a dashed black line, with the 0.3 dex scatter shaded in gray. In the quiescent population, $\DeltaL$ is strongly negative (the average across the quiescent galaxies is $\langle \DeltaL \rangle_{\rm{Q}} = -1.34^{+0.41}_{-0.51}$ dex) while $\DeltaS$ is much less negative ($\langle \DeltaS \rangle_{\rm{Q}} = -0.29^{+0.31}_{-0.43}$ dex), leading to largely negative values of $|\DeltaS| - |\DeltaL|$ as well (on average, $\langle |\DeltaS| - |\DeltaL| \rangle_{\rm{Q}} = -1.02^{+0.78}_{-0.57}$ dex). This informs us that, as we might expect, a quiescent galaxy's position off of the SFMS is predominantly determined by long-timescale SFH variations. In other words, quenching in these massive galaxies is likely a gradual process that takes place over long ($\gtrsim$ Gyr) timescales.

Perhaps more interestingly, we find that on the SFMS, $\DeltaL$ and $\DeltaS$ --- and thus, by construction, $|\DeltaS| - |\DeltaL|$ --- are all consistent with zero. Across the star-forming population, we find median values of $\langle \DeltaL \rangle_{\rm{SF}} = -0.05^{+0.27}_{-0.29}$ dex, $\langle \DeltaS \rangle_{\rm{SF}} = 0.12^{+0.24}_{-0.27}$ dex, and $\langle |\DeltaS| - |\DeltaL| \rangle_{\rm{SF}} = -0.01^{+0.21}_{-0.23}$ dex. At a glance, this seems to imply that on average, star-forming galaxies have remained around the SFMS for at least the last $\sim 1$ Gyr and that long- and short-term movements in a star-forming galaxy's SFH contribute equally to its current position on the main sequence. However, it is not immediately clear from this figure alone whether $\DeltaL$, $\DeltaS$, and $|\DeltaS| - |\DeltaL|$ are all $\sim$ 0 dex because both $\DeltaL$ and $\DeltaL$ are close to zero across the entire SFMS, or whether it is because, e.g. $\DeltaL$ and $\DeltaS$ are both similarly negative in galaxies below the SFMS and similarly positive for those above. That is to say, do galaxies oscillate around \textit{the same} SFMS, or do they fluctuate around \textit{different} median relations? We further investigate this below.

\begin{figure*}
    \centering
    \includegraphics[width=0.95\textwidth]{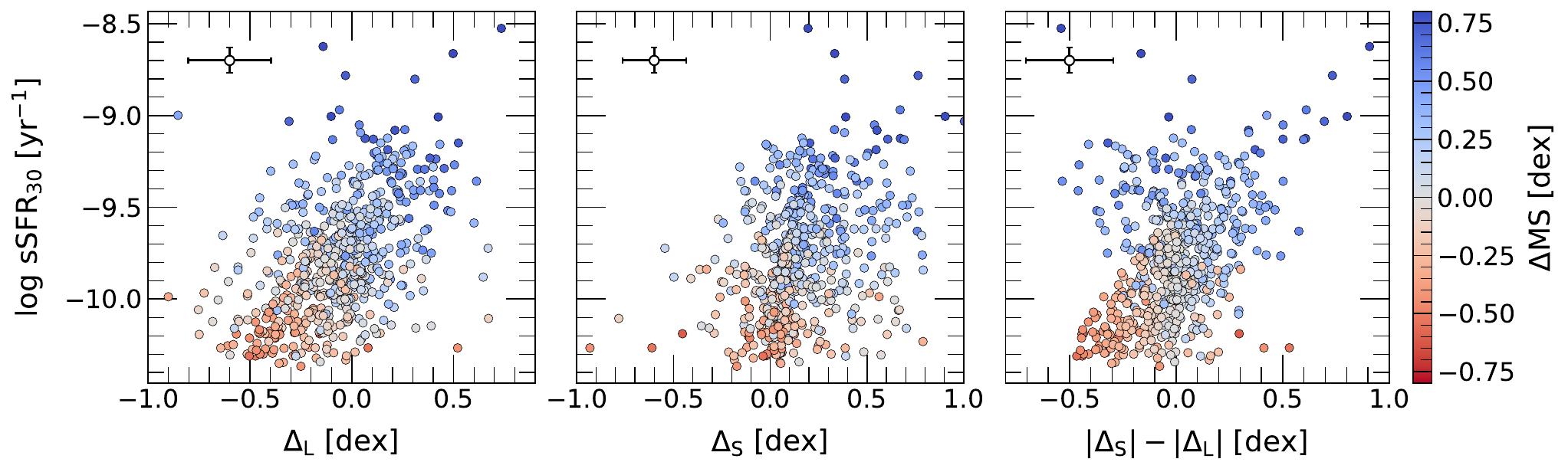}
    \caption{log sSFR versus (from left to right) $\Delta_{\rm{L}}$, $\Delta_{\rm{S}}$, and $|\Delta_{\rm{S}}| - |\Delta_{\rm{L}}|$ for the star-forming galaxies in our sample. The points are further color-coded by $\DeltaMS$ in each panel. $\DeltaL$ experiences clear a diverging behavior, i.e., $\DeltaL$ is more negative below the SFMS and more positive above. This potentially indicates the presence of different median relations in the star-forming population. $\DeltaS$ is roughly scattered around 0 dex for the entire star-forming population; however, there is also a weak increase from below to above the SFMS. Additionally, below the SFMS, $|\DeltaS| - |\DeltaL| < 0$ dex, implying that the lower levels of star-formation present in these systems are predominantly due to long-timescale processes and could point to quenching in action. Above the SFMS, $|\DeltaS| - |\DeltaL|$ spans values between $-0.5 - 0.8$ dex, highlighting the manifold processes that regulate star formation.}
    \label{fig:timescales_sf}
\end{figure*}

Figure \ref{fig:timescales_sf} is a reconfigured version of Figure \ref{fig:timescales}, this time zoomed-in on just the star-forming population. 
We plot log sSFR versus (from left to right) $\DeltaL$, $\DeltaS$, and $|\DeltaS| - |\DeltaL|$. In each panel, the points are color-coded by $\DeltaMS$ to further highlight trends with star-formation activity. Characteristic error bars are also shown in the upper left corners of each panel.

Focusing on the left-hand panel of Figure \ref{fig:timescales_sf}, we see that there is a clear relationship between $\DeltaL$ and level of star formation. As log sSFR (or equivalently, $\DeltaMS$) increases, so does $\DeltaL$. Additionally, while galaxies above the SFMS ridgeline ($\DeltaMS = 0$) tend to have positive values of $\DeltaL$, those below tend to have negative values of $\DeltaL$. The median value of $\DeltaL$ in galaxies which are more than 0.3 dex above the SFMS is $0.19^{+0.28}_{-0.23}$ dex; in galaxies more than 0.3 dex below the SFMS, this value is $-0.39^{+0.14}_{-0.07}$ dex; and in galaxies within $\pm 0.3$ dex of the main sequence ridgeline, the median is $-0.07^{+0.17}_{-0.22}$ dex.

There is a much weaker trend between $\DeltaS$ and star-formation activity (middle panel of Figure \ref{fig:timescales_sf}). While the majority of galaxies have $\DeltaS$ values around 0 dex, there is a mild increase in $\DeltaS$ with log sSFR and $\DeltaMS$. We find a median $\DeltaS$ of $0.24^{+0.29}_{-0.18}$ dex in galaxies that are more than 0.3 dex above the SFMS ridgeline; $0.09^{+0.05}_{-0.15}$ dex for those more than 0.3 dex below the SFMS; and $0.02^{+0.16}_{-0.12}$ dex for those within the $\pm 0.3$ dex envelope. It is interesting to note that across the main sequence, $\DeltaS$ tends to remain positive, while there is a clear divergence in the sign of $\DeltaL$ in systems above and below the SFMS. 

The combination of these two components translate into $|\DeltaS| - |\DeltaL|$ (right-hand panel of Figure \ref{fig:timescales_sf}) being, on average, more positive in systems on the SFMS and more negative below the SFMS. The median $|\DeltaS| - |\DeltaL|$ in galaxies further than 0.3 dex below the SFMS is $-0.31^{+0.19}_{-0.08}$ dex and $-0.01^{+0.13}_{-0.17}$ dex in galaxies within 0.3 dex of the main sequence. However, the picture becomes more complicated looking at systems further than $0.3$ dex above the SFMS. As seen in the right-most panel of Figure \ref{fig:timescales_sf}, systems with very high star-formation span a wide range, with $-0.5~\mathrm{dex} \lesssim |\DeltaS| - |\DeltaL| \lesssim 0.8~\mathrm{dex}$. Unlike galaxies on and below the SFMS, there is no clean relationship between $|\DeltaS| - |\DeltaL|$ and sSFR in highly star-forming galaxies.

By breaking down the short- and long-term behavior of star-forming galaxies significantly above, on, and significantly below the SFMS, we can see that these three groups are host to qualitatively different SFHs. The fact that $\DeltaL \sim 0.2$ dex in systems more than 0.3 dex above the main sequence ridgeline tells us that they have been above the main sequence for the past $\sim$ Gyr. In parallel, $\DeltaL \sim -0.4$ dex in galaxies more than 0.3 dex below the SFMS, meaning they tend to have been below the SFMS over the last Gyr. All the while, within the $\pm 0.3$ dex envelope around the SFMS, $\DeltaL \sim 0$ --- galaxies around the main sequence ridgeline have historically been around the ridgeline for a Gyr. Together, these pieces of information imply that individual galaxies follow disparate median paths throughout their star-forming lifetimes.

Furthermore, we find that $\DeltaS \sim 0$ dex in galaxies significantly below the main sequence, resulting in $|\DeltaS| - |\DeltaL| \sim -0.3$ dex. This suggests that short-term processes currently play little role in determining their star-formation activity. It could be the case that these systems have slowly been winding down their star formation and are now beginning to quench, which would explain their long-term positions below the main sequence with relatively little up-regulation from, e.g., gas compaction events.

On the other hand, $\DeltaS \sim 0.2$ dex in systems $> 0.3$ dex above the SFMS. This means that galaxies significantly above the SFMS have not only spent a sizeable fraction of their lives above the SFMS, but short-term processes also tend to boost their SFRs. However, there is a marked lack of consensus between the relative strengths of $\DeltaS$ and $\DeltaL$ in these highly star-forming systems  --- there is over a decade of spread in the values of $|\DeltaS| - |\DeltaL|$. This may be a sign of different star-formation triggering mechanisms at play. It is possible that systems with $|\DeltaS| - |\DeltaL| > 0$ dex are experiencing starbursts caused by a recent dynamical interaction, while those with $|\DeltaS| - |\DeltaL| < 0$ dex have high SFRs sustained by long-timescale, steady gas inflows. However, more analysis is necessary to reach a conclusion. 

For galaxies within the $\pm 0.3$ dex SFMS envelope, $\DeltaS$ and $\DeltaL$ are comparable in magnitude ($|\DeltaS| - |\DeltaL| \sim 0$ dex). Moreover, $\DeltaL$ and $\DeltaS$ are also both individually $\sim 0$ dex, which is an interesting statement about the self-regulation of star formation. In particular, it may imply that in the most ``typical'' star-forming galaxies, the star-formation process is regulated remarkably well such that significant deviations from the SFMS on both short and long timescales are rare. It also means that these galaxies' positions relative to the SFMS result from equal parts long- and short-timescale star-formation activity.

We can examine short-term movement on the SFMS a different way by calculating the recent SFR gradient (or change in SFR) of the galaxies in our sample. Following \citet{Ciesla2023}, we define the SFR gradient as the angle from horizontal created by linking the position of the galaxy $\Delta t$ years ago to its current position in SFH space. In other words,
\begin{equation}
    \nabla \mathrm{SFR}_{\Delta t} = \arctan \Bigg{(} \frac{\log \mathrm{SFR}_{t_0} - \log \mathrm{SFR}_{t_0 + \Delta t}}{\Delta t} \Bigg{)},
\end{equation}
where $t_0$ indicates the time at observation. $\nabla \rm{SFR} > 0$ indicates an increasing SFR, while $\nabla \rm{SFR} < 0$ indicates a decreasing SFR, and the magnitude of the angle represents the steepness of the change in SFR. 

\begin{figure}
    \centering
    \includegraphics[width=0.48\textwidth]{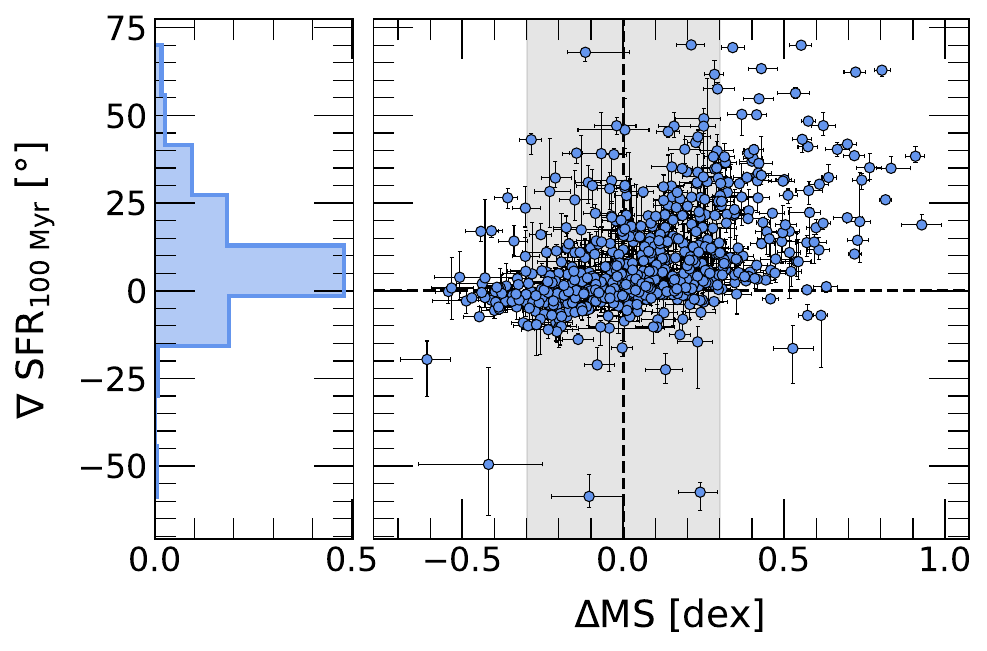}
    \caption{SFR gradient (i.e., change in SFR) over the last 100 Myr ($\nabla \rm{SFR_{100~Myr}}$) versus distance from the SFMS ($\DeltaMS$) for the star-forming galaxies in our sample. We display the marginal distribution of of $\nabla \rm{SFR_{100~Myr}}$ in the left-hand panel. We find that while on average galaxies on the SFMS have fairly constant SFRs over the last 100 Myr, many systems also have experienced recent increases (and some decreases) in their star-formation activity, indicating that galaxies do move about the SFMS on timescales significantly shorter than the age of the universe.}
    \label{fig:sfr gradient delMS}
\end{figure}

In the main panel of Figure \ref{fig:sfr gradient delMS}, we plot SFR gradient over $\Delta t = 100$ Myr ($\nabla$SFR$_{\mathrm{100~Myr}}$) as a function of current distance from the SFMS (in logarithmic space; $\DeltaMS$). The marginal distribution of $\nabla$SFR$_{\mathrm{100 Myr}}$ is shown in the left-hand panel. We find that the typical $\nabla$SFR$_{\mathrm{100~Myr}}$ value for our star-forming population is actually consistent with zero ($4.47 \degree^{+16.78 \degree}_{-6.60 \degree}$), meaning that on average, galaxies on the SFMS have remained on the SFMS over 100 Myr timescales. 

However, from the range of $\nabla$ SFR$_{\mathrm{100~Myr}}$ values seen in Figure \ref{fig:sfr gradient delMS}, it is clear that these galaxies show appreciable movement relative to the SFMS in recent times. In fact, $\sim 15$\% of the star-forming galaxies in our sample have $|\nabla \rm{SFR_{100~Myr}}| > 25\degree$, translating to a $\gtrsim10$\% increase in their SFR over the last 100 Myr. There is a noticeable tail towards large values of $\nabla \rm{SFR_{100~Myr}}$, meaning a significant portion of the star-forming population have recently increased their SFRs. This echoes the positive skewness of $\DeltaS$, as seen in Figure \ref{fig:timescales_sf}. We also see a weak positive correlation between $\nabla$ SFR$_{\mathrm{100~Myr}}$ and $\DeltaMS$, implying that galaxies above the SFMS have experienced recent increases in their SFRs (and vice versa for galaxies below the SFMS), in line with the observed trend in $\DeltaS$ in Figure \ref{fig:timescales_sf}. This confirms that, despite their general ``stay the course'' tendencies, galaxies do also move about the SFMS on rather short timescales, which is consistent with previous observational studies \citep[e.g.,][]{Ciesla2017, Tacchella2022Halo7D}, as well as simulations \citep[e.g.,][]{Tacchella2016, Iyer2020}. 

Altogether, this paints an interesting picture. While on average, star-forming galaxies obey the nominal SFR$-\Mstar$ relation, individual galaxies may actually oscillate around \textit{different} median relations. Short-term variations also exist on top of these long-timescale paths, boosting the short-timescale scatter in the SFMS.

\section{Discussion}
\label{sec:discussion}

In this section, we answer the hard-hitting questions: \textit{How should we think about the star-forming main sequence?} and \textit{So, what next?} We discuss the interpretation of the SFMS in light of our analysis in Section \ref{sec:interpreting the SFMS} and avenues for future work in Section \ref{sec:future work}.

\subsection{Interpreting the SFMS}
\label{sec:interpreting the SFMS}

The primary objective of this work was to distinguish between two different explanations of the star-forming main sequence --- one where its scatter results from individual galaxies oscillating around the same median relation as they evolve, and another where this relation reflects a population average, with its scatter representing the diversity of SFHs in the galaxy population. So where do we stand?

In Section \ref{sec:what's in a scatter}, we looked at the behavior of the scatter of the SFMS as a function of timescale over which the measured SFRs are averaged. We found that $\sigmaMS$ is largest ($\sim 0.3$ dex) at short-timescales and decreases with averaging timescale. However, $\sigmaMS$ remains non-negligible ($\sim 0.15 - 0.25$ dex) out to 2 Gyr at all masses. This highlights the importance of long-timescale ($\gtrsim 1$ Gyr) variations to the observed scatter in the SFR$-\Mstar$ plane. This is in line with previous findings that the SFHs of massive, star-forming galaxies retain their ``memory'' over a large fraction of the Hubble time ($\sim3$ Gyr; \citealt{Chauke2018}). Additionally, while the steepness of the decline in $\sigmaMS$ with averaging timescale depends on stellar mass, this overall trend indicates that galaxy SFHs experience short-term ($\lesssim 100$ Myr) fluctuations as well.

In Section \ref{sec:what's in an offset}, we investigated galaxies' median positions relative the SFMS on long (1 Gyr) and short ($\lesssim$ 30 Myr) timescales. We observed that galaxies above the 0.3 dex SFMS envelope tend to have been above the SFMS for at least $\sim 1$ Gyr (and likewise for galaxies below the SFMS), again pointing to the importance of the long-timescale behavior of SFHs. Furthermore, it provides evidence that galaxies may oscillate around \textit{different} median SFR$-\Mstar$ relations. However, galaxies' long-term star-formation activity cannot entirely explain their current positions on the main sequence. On top of their median SFH tracks, galaxies also fluctuate in their SFRs over tens to hundreds of Myr timescales.

\citet{Matthee2019} present a similar finding in their analysis of star-forming galaxies in the EAGLE simulations. They demonstrate that galaxies above the main sequence at $z= 0.1$ have median SFRs that place them above the main sequence for $\sim 10$ Gyr, but also experience fluctuations of $\sim 0.2$ dex on shorter timescales ($\lesssim$ 2 Gyr). They isolate the origin of the long timescale fluctuations as results of differences in halo mass and formation time (i.e., assembly bias). Conversely, the short timescale fluctuations do not simply align with changes in halo accretion, but are more likely associated to the self-regulation of star formation via feedback processes.

While we cannot distinguish exactly what processes establish SFH variations over different timescales in this work, it is plausible that short-term variations in galaxy SFHs arise from the complex interplay between gas inflow, outflow, and consumption \citep[e.g.,][]{Tacchella2016, Jain2024}, while long-term variations reflect the environments in which the galaxies live \citep[e.g.,][]{Coil2017, Matthee2019, Berti2021}. Thus, it may be that over the long-term, individual galaxies follow different SFH tracks set by their parent halo properties, but experience perturbations around this track on shorter timescales. In other words, the SFMS relation reflects the average paths galaxies follow in the SFR$-\Mstar$ plane, and its scatter arises from both the heterogeneity of these paths, as well as the short-timescale variations individual galaxies experience on top of them as a result of feedback processes.

\subsection{Limits, caveats, and future work}
\label{sec:future work}

While the high resolution of the LEGA-C spectra have facilitated the in-depth analysis of star-formation histories presented here, there are still, of course, some limitations. The galaxies studied in this work are confined to massive systems above $\sim 10^{10}~\Msun$ and span about a decade and a half in stellar mass. Additionally, we are only mass-complete above $\sim 10^{10.9}~\Msun$ for the full $0.6 \leq z \leq 1.0$ redshift range that we probe. Thus, any statements about mass-dependence in the scatter of the main sequence are restricted to quite a narrow mass range, and trends should be interpreted with a grain of salt.

Furthermore, our analysis of SFHs on long timescales is limited by how far back in a galaxy's history we are able to derive meaningful constraints. Because older stellar populations will always be outshone by younger ones (an effect known as ``outshining''), it is inherently more difficult to estimate the SFHs of galaxies at large lookback times. On the other end, our analysis of short-timescale SFH fluctuations is limited by our time resolution of SFHs. Because we cannot bin our SFHs infinitely finely (or even as finely as a simulation snapshot cadence), we will always miss variations on the shortest timescales.

It is also unfortunate that the current state of SED-fitting tools and the stochastic SFH prior does not allow for constraints on the underlying PSD from which a galaxy population's SFHs are ``drawn'' at the modelling step. However, as discussed in Section \ref{sec:what's in a scatter}, it \textit{is} possible to do so using the scatter of the main sequence measured over different SFR averaging timescales. We have not attempted this in our analysis, but it is definitely worth doing in the future.

Furthermore, investigations into trends in stellar and gas-phase metallicity, as well as dust properties, as a function of star-formation activity are outside the scope of this work. However, it would be interesting to see if we can glean any information about, e.g., what processes are at work in galaxies far above the SFMS. Is there evidence of inflowing pristine gas replenishing gas reservoirs in these systems (thereby lowering the gas-phase metallicity relative to the stellar metallicity) and causing starbursts?

There are also a lot of questions about galaxies off the SFMS that can (and should!) be asked of this dataset that we did not explore in this work. In particular, there is a lot that can be learned about the quenching process (in massive galaxies) through SFH timescale analyses similar to the ones applied to star-forming galaxies in this paper. How long do galaxies spend in the green valley (the transition zone between star-forming and quenched systems)? Does quenching happen quickly or slowly? On what timescales did quiescent galaxies form their stars? What causes rejuvenation? Addressing these questions can build our knowledge of how, when, and why galaxies move off the main sequence, and definitely warrants careful thought and analysis in the future.

Looking further out, surveys like MOONRISE \citep{MOONRISE2020}, the main Guaranteed Time Observation MOONS (the Multi-Object Optical and Near-infrared Spectrograph) extragalactic survey, will be able to provide high-resolution spectra of hundreds of thousands of galaxies at cosmic noon ($z \sim 1-2.5$). This will allow us to characterize the environments in which these adolescent galaxies lived and evolved in, and understand the relationship between galaxy SFHs and environment at the epoch in which the star-formation rate of the Universe peaked. At the same time, current (and upcoming) JWST surveys are giving us an unprecedented window into star formation in the early universe. As we continue to stockpile spectra for these high-$z$ systems, we will also be able to understand how star-formation variability evolves throughout cosmic time.

\section{Conclusions}
\label{sec:conclusions}

In this work, we jointly model the spectra (from the LEGA-C survey) and photometry (from the COSMOS2020 and Super-deblended catalogs) of 1928 massive galaxies at $z = 0.6-1.0$ using the SED-fitting code {\typewriter Prospector} with the model parameters described in Table \ref{tab:priors}. By applying the stochastic SFH prior \citep{Wan2024}, we are able to obtain high-fidelity estimates of galaxy SFHs, allowing us to analyze the origin of the SFMS.

We first present the reconstructed SFHs of all the galaxies in our sample, categorized by evolutionary regime: star-forming, transitioning, and quiescent. These SFHs highlight the diversity in the pathways through which galaxies evolve while also revealing some overall patterns. Star-forming galaxies show continuous star formation over long periods, while transitioning galaxies show an initial peak in star formation that declines over time. Massive quiescent galaxies, on the other hand, formed most of their stars early and have very little to no star formation at present 
A small number of galaxies show evidence of rejuvenation, where star formation reignites after a period of quiescence, though this is rare and contributes minimally to overall stellar mass. These qualitative SFH trends can be quantified by the mass-weighted ages of each galaxy population. Star-forming galaxies tend to be the youngest, with median ages of around 2.8 Gyr, while quiescent galaxies are older, with median ages of around 4.5 Gyr. Transitioning galaxies fall in between. 

Within the star-forming population, we observe a weak age gradient across the SFMS, where more massive galaxies tend to be older than their lower-mass counterparts. This implies that there is a long-term correlation in determining whether galaxies are above or below the SFMS ridgeline. We find additional evidence for this by examining the scatter of the SFMS, as well as individual galaxies' offsets from the SFMS.

We determine that the scatter in the SFRs of star-forming galaxies with stellar masses $\gtrsim 10^{10}~\Msun$ is $\sim 0.3$ dex, decreasing very slightly with stellar masses. Measuring this scatter as a function of the timescale over which SFRs are averaged reveals a decline with averaging timescale. Specifically, the scatter decreases from $\sim 0.3$ dex at a timescale of 10 Myr to $\sim 0.15$ dex in galaxies with stellar masses in the range $10^{10.25} < \Mstar/\Msun < 10^{10.75}$; $\sim 0.2$ dex in galaxies with stellar masses in the range $10^{10.75} < \Mstar/\Msun < 10^{11.0}$; and $\sim 0.25$ dex in galaxies with stellar masses in the range $10^{11} < \Mstar/\Msun < 10^{11.75}$ at a timescale of 2 Gyr. This decrease occurs more rapidly at lower stellar masses, indicating that lower-mass galaxies tend to experience more variations in their SFHs over short timescales. However, in all cases, the scatter measured at an averaging timescale of 2 Gyr remains non-negligible, highlighting the importance of Gyr-length SFR fluctuations to the scatter observed in the SFR$-\Mstar$ plane.

We break down the offsets of galaxies from the SFMS into long-term ($\sim 1$ Gyr; $\DeltaL$) and short-term ($\lesssim 30$ Myr; $\DeltaS$) components. We find that galaxies that are currently above (below) the SFMS have tend to have remained above (below) the SFMS over the last Gyr. This provides evidence that galaxies may follow \textit{different} median SFR$-\Mstar$ relations that, over the population, simply average out to SFMS relation we observe. On top of these long-term median trends, however, galaxies also oscillate in their SFRs by $\sim 0.1-0.2$ dex over Myr-timescales. 

As a whole, our results illustrate that variations in both the long- and short-timescale behavior of galaxies' SFHs are represented in the scatter of the main sequence. The SFMS is an evolutionary sequence in the sense that galaxies do not simply pass through the main sequence once on their various paths through the SFR$-\Mstar$ space. However, individual galaxies may move along distinct median SFH tracks that are set by their environments and parent halo properties, oscillating around those tracks on shorter timescales due to the interplay between gas inflow, outflow and consumption associated with the regulation star-formation.

\section*{Acknowledgements}
JTW was supported by the Churchill Scholarship through the Winston Churchill Foundation. JTW acknowledges the examiners of her MPhil viva for providing helpful feedback that improved this work. This work was performed using resources provided by the Cambridge Service for Data Driven Discovery (CSD3) operated by the University of Cambridge Research Computing Service (\url{www.csd3.cam.ac.uk}), provided by Dell EMC and Intel using Tier-2 funding from the Engineering and Physical Sciences Research Council (capital grant EP/T022159/1), and DiRAC funding from the Science and Technology Facilities Council (\url{www.dirac.ac.uk}).

\textit{Software}: {\typewriter astropy} \citep{Astropy2022}, {\typewriter dynesty} \citep{Speagle2020}, {\typewriter GP-SFH} \citep{Iyer2024}, {\typewriter matplotlib} \citep{Matplotlib2007}, {\typewriter numpy} \citep{Numpy2020}, {\typewriter Prospector} \citep{Johnson2021}.

\section*{Data Availability}
The LEGA-C DR3 data used in this analysis is publicly available at \url{https://users.ugent.be/~avdrwel/research.html#legac}. The best-fitting SED parameters for the galaxies in this analysis are available as supplementary material.


\bibliographystyle{mnras}
\bibliography{SFH_paper}

\appendix

\section{Comparing stellar masses against LEGA-C DR3}
\label{app:comparison}

A number of previous works have presented SED-fitting results of LEGA-C galaxies \citep[e.g.][]{Cappellari2023, Kaushal2024, Steel2024, Nersesian2025}. While performing a census of various models' estimates of key galaxy parameters and investigating their consistency (or potential lack thereof) is a very interesting and informative exercise, it is outside the scope of this work. Here, as a simple case-study, we provide a direct comparison between the stellar mass (log M$_*$) 
values derived from the {\typewriter Prospector} fits used in this work against the stellar masses in the LEGA-C DR3 catalog \citep{LEGAC2021}, as well as the best-fit values from \citet{Cappellari2023}.

The stellar masses in this work were obtained from {\typewriter Prospector} joint fits to the LEGA-C DR3 spectra and COSMOS2020 \citep{COSMOS2020} photometry using the model described in Section \ref{sec:model and priors}. The LEGA-C DR3 stellar masses were determined by fitting only the \textit{BVrizYJ} photometry using {\typewriter Prospector} (see \citealt{LEGAC2021} for details). And the \citep{Cappellari2023} mass estimates come from {\typewriter pPXF} \citep{Cappellari2017} fits to the LEGA-C DR3 spectra and photometry from the COSMOS2020 catalog, as well as the UltraVISTA/COSMOS catalog \citep{Muzzin2013}. We have adjusted both the LEGA-C DR3 and \citet{Cappellari2023} stellar masses to account for the fact that the former assumed a \citet{Chabrier2003} IMF and the latter model assumed a \citet{Salpeter1955} IMF, while we use a \citet{Kroupa2001} IMF. This is accomplished with a 0.05 dex and 0.18 dex shift in the stellar masses, respectively \citep{Madau2014}.

\begin{figure}
    \centering
        \begin{subfigure}[b]{0.39\textwidth}
        \centering
        \includegraphics[width=\textwidth]{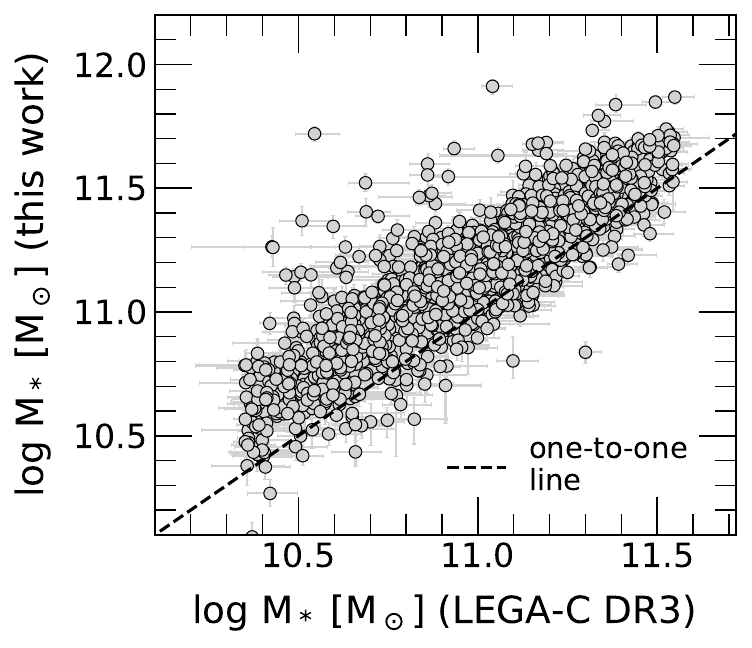}
    \end{subfigure}
    \hfill
    \begin{subfigure}[b]{0.39\textwidth}
        \centering
        \includegraphics[width=\textwidth]{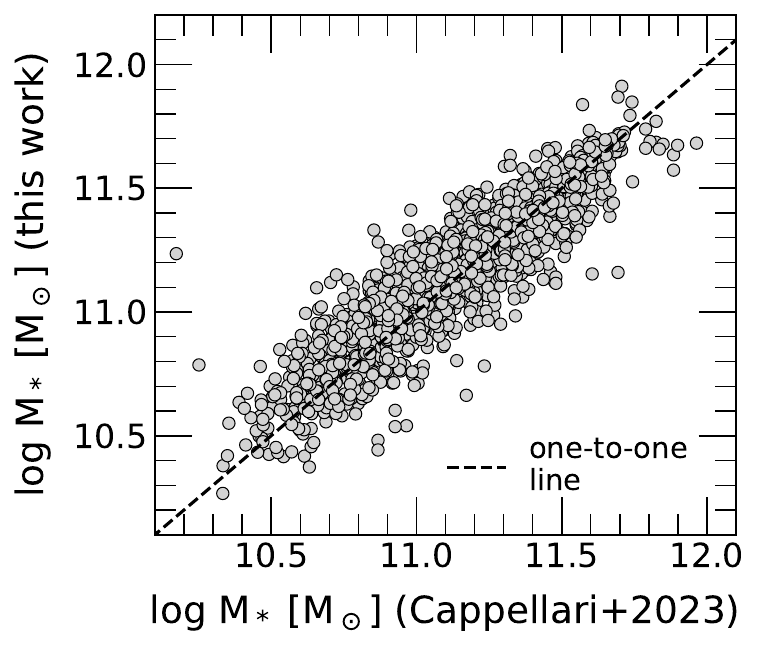}
    \end{subfigure}
    \caption{Comparison of the stellar masses from this work with the LEGA-C DR3 \citep{LEGAC2021} catalog (top) and \citet{Cappellari2023} best-fit (bottom) stellar masses. We find that our median stellar mass estimates are biased high by $\sim 0.15$ dex relative to the LEGA-C DR3 values; this difference is likely due to the fact that we jointly model the LEGA-C spectra and photometry covering the near-UV to the IR, while \citet{LEGAC2021} only model the \textit{BVrizYJ} photometric bands. On the other hand, the stellar masses estimated in \citet{Cappellari2023} which, like this work, uses both the LEGA-C DR3 spectra and photometry from the COSMOS2020 catalog \citep{COSMOS2020}, are largely consistent with our estimated stellar masses.}
    \label{fig:param_comparison}
\end{figure}

The top panel of Figure \ref{fig:param_comparison} plots the stellar masses derived in this work against the LEGA-C DR3 values. Overall, we find that there is a $\sim 0.15$ dex systematic offset in stellar mass. This is likely a due to the fact that \citet{LEGAC2021} performed SED-fitting of rest-optical photometry only (\textit{BVrizYJ} photometric bands), which could lead to an underestimation of stellar masses if dust is not properly accounted for.

The bottom panel of Figure \ref{fig:param_comparison} shows the comparison with \citet{Cappellari2023}. Despite using different models, we are actually very consistent with the \citet{Cappellari2023} stellar masses. Similar to our work, \citet{Cappellari2023} stellar masses were derived from joint fits to the LEGA-C survey spectra and photometry from the COSMOS2020 catalog \citep{COSMOS2020}. This provides further evidence that the offset between our stellar masses and the LEGA-C DR3 values are due to differences in the data being modelled.

\section{Dust and metallicity}
\label{app:dust and met}

\begin{figure}
    \centering
        \begin{subfigure}[b]{0.48\textwidth}
        \centering
        \includegraphics[width=\textwidth]{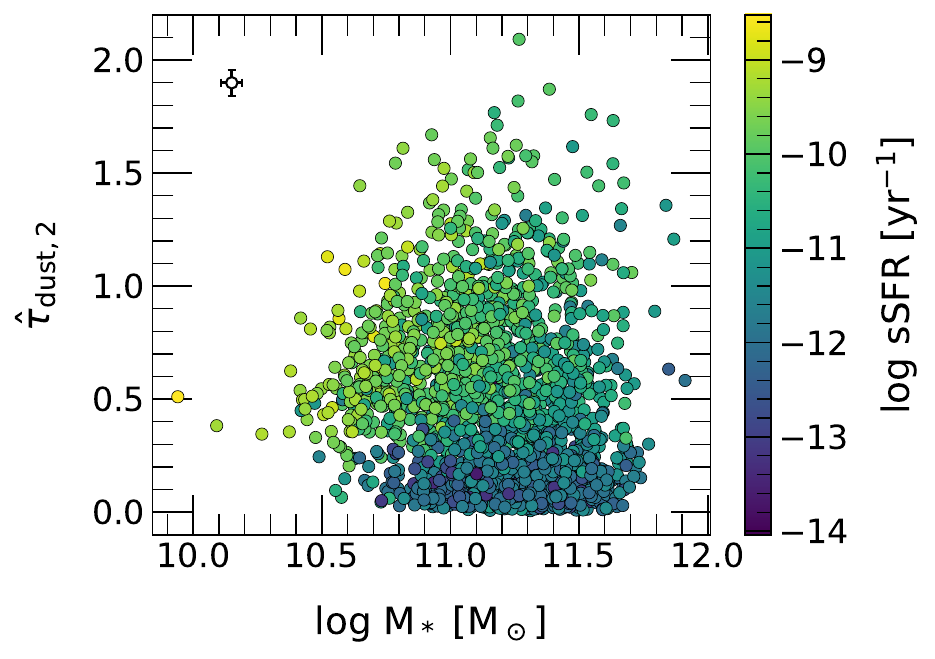}
    \end{subfigure}
    \hfill
    \begin{subfigure}[b]{0.48\textwidth}
        \centering
        \includegraphics[width=\textwidth]{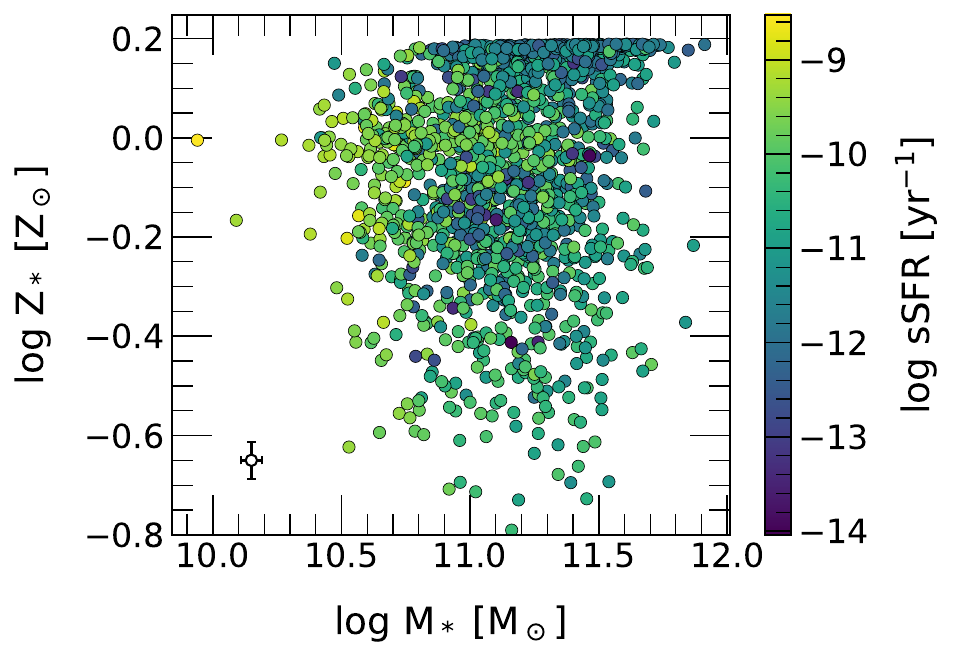}
    \end{subfigure}
    \hfill
    \begin{subfigure}[b]{0.48\textwidth}
        \centering
        \includegraphics[width=\textwidth]{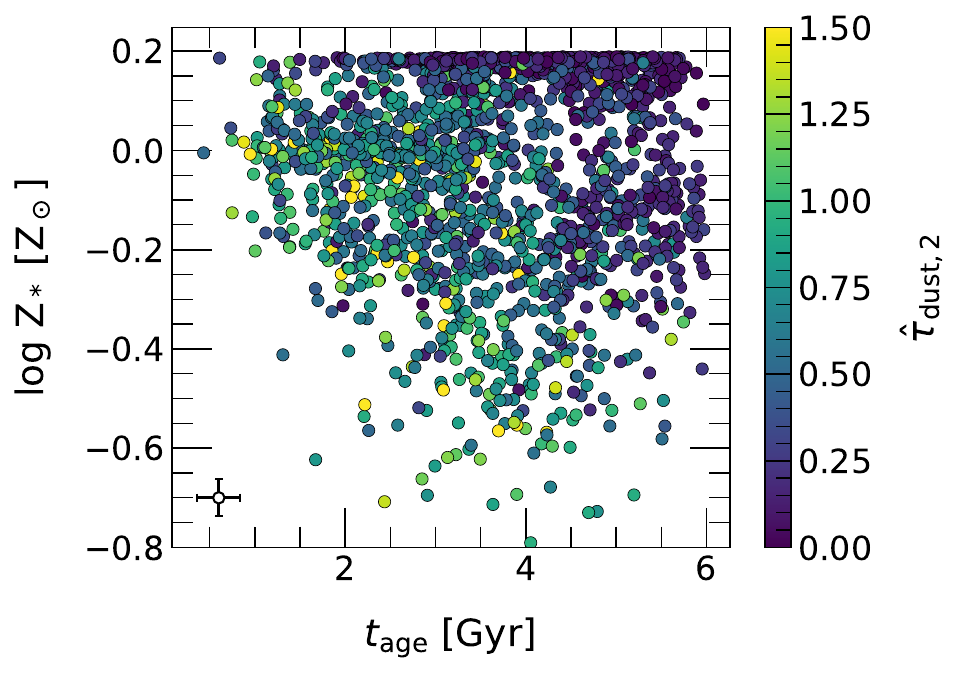}
    \end{subfigure}
    \caption{\textit{Top:} Diffuse dust optical depth ($\hat{\tau}_{\rm{dust,2}}$) as a function of stellar mass (log M$_*$), color-coded by specific star-formation rate (log sSFR). Star-forming galaxies (i.e. galaxies with high sSFRs) tend to be dustier than quiescent galaxies (i.e. low sSFRs), and there is a weak positive correlation between dust attenuation and mass in the star-forming systems.
    \textit{Middle:} Stellar metallicity (log Z$_*$) as a function of stellar mass, color-coded by specific star-formation rate. Star-forming galaxies have stellar metallicities consistent with solar, while quiescent galaxies tend to be more enriched. 
    \textit{Bottom:} Stellar metallicity as a function of mass-weighted age ($\tage$), color-coded by diffuse dust optical depth. We see that early-forming galaxies (i.e. older $\tage$ values) tend to be more metal-poor than late-forming ones.
    Characteristic error bars are shown in each panel.}
    \label{fig:dust and met}
\end{figure}

We present here a brief overview of the diffuse dust optical depths ($\hat{\tau}_{\rm{dust,2}}$) and stellar metallicities (log Z$_*$) measured for the galaxies in our sample (Figure \ref{fig:dust and met}).

The top panel of Figure \ref{fig:dust and met} shows $\hat{\tau}_{\rm{dust,2}}$ as a function of stellar mass (log M$_*$), color-coded by specific star-formation rate (log sSFR). We see that star-forming galaxies (i.e. galaxies with high sSFRs) experience stronger dust attenuation than quiescent galaxies (i.e. galaxies with low sSFRs). Additionally, in the star-forming systems, there tends to be an increase in dust attenuation with stellar mass. Both of these features are qualitatively consistent with well-known trends found by the numerous studies of galaxy evolution and dust \citep[e.g.][]{Kong2004, Johnson2007, Battisti2016}.

The middle panel of Figure \ref{fig:dust and met} plots log Z$_*$ as a function of stellar mass (log M$_*$), color-coded by specific star-formation rate (log sSFR). We find that star-forming galaxies are generally consistent with solar metallicity. Quiescent galaxies, on the other hand, tend to be more metal-enriched than their star-forming counterparts by up to $\sim0.2$ dex, consistent with previous findings from, e.g. \citet{Peng2015} and \citet{Trussler2020}. Additionally, in the quiescent population, lower-mass galaxies span a range of metallicities, while the higher-mass galaxies cluster at high metallicites, qualitatively similar to \citet{Bevacqua2024}.

Lastly, the bottom panel of Figure \ref{fig:dust and met} plots log Z$_*$ versus galaxy mass-weighted age ($\tage$), color-coded by $\hat{\tau}_{\rm{dust,2}}$. In the star-forming population in particular (the lighter-colored, ``dustier'' points), stellar metallicity has a strong dependence on age --- galaxies that formed earlier (i.e. have older $\tage$ values) tend to be much more metal-poor than those that formed more recently. This is consistent with the understanding that chemical enrichment progresses over cosmic time, and thus galaxies forming earlier in the Universe's history have a lower metal content \citep[e.g.][]{Yuan2013, Langeroodi2023}. It is also worth noting that {\typewriter Prospector} assumes a single scaled-solar abundance pattern for a galaxy, meaning that alpha enhancement is not modelled. This can lead to an overestimation of metallicity \citep{Beverage2021, Bevacqua2023, Beverage2024, Park2024, Turner2025}, which could explain the galaxies hitting the upper limit in log Z$_*$, and in particular the early-forming systems.


\bsp	
\label{lastpage}
\end{document}